\documentclass[a4paper, 12pt]{article}

\usepackage{indentfirst}
\usepackage{graphicx}
\usepackage{overpic}
\usepackage{array}
\usepackage{color}
\usepackage{multirow}
\usepackage{float}
\usepackage{subfigure}
\usepackage{amsmath,amssymb, amsthm}
\usepackage[mathscr]{euscript}
\usepackage{latexsym,bm}
\usepackage{cite}
\usepackage{booktabs}
\usepackage{appendix}
\usepackage{bbm}

\newcommand\comment[1]{}

{\theoremstyle{remark} }

\def\mi{\mathbbm{i}}
\def\me{\mathbbm{e}}
\def\D{\text{d}}

\begin{document}

\bibliographystyle{unsrt}

\title{A higher-order accurate operator splitting spectral method for the Wigner-Poisson system}
\author{Zhenzhu Chen\footnotemark[2] $^,$\footnotemark[3]
\and Haiyan Jiang \footnotemark[4]
\and Sihong Shao\footnotemark[3] $^,$\footnotemark[1]}
\renewcommand{\thefootnote}{\fnsymbol{footnote}}
\footnotetext[2]{Institute of Applied Physics and Computational Mathematics, Beijing 100094, China.}
\footnotetext[3]{CAPT, LMAM and School of Mathematical Sciences, Peking University, Beijing 100871, China.}
\footnotetext[4]{School of Mathematics and Statistics, Beijing Institute of Technology, Beijing 100081, China.}
\footnotetext[1]{To
whom correspondence should be addressed. Email:
\texttt{sihong@math.pku.edu.cn}}
\date{\today}
\maketitle

\begin{abstract}

 An accurate description of 2-D quantum transport in a double-gate metal oxide semiconductor filed effect transistor (dgMOSFET) requires a
 high-resolution solver to a coupled system of the 4-D Wigner equation and 2-D Poisson equation. In this paper, we propose an operator splitting spectral method to evolve such Wigner-Poisson (WP) system in 4-D phase space with high accuracy. After an operator splitting of the Wigner equation, the resulting two sub-equations can be solved analytically with spectral approximation in phase space. Meanwhile, we adopt a Chebyshev spectral method to solve the Poisson equation. Spectral convergence in phase space and a fourth-order accuracy in time are both numerically verified. Finally, we apply the proposed solver into simulating dgMOSFET,
 develop the steady states from long-time simulations and obtain numerically converged current-voltage (I-V) curves.
 
 \vspace*{4mm}
\noindent {\bf Keywords:}
Wigner-Poisson system;
operator splitting;
spectral method;
MOSFET;
I-V curve;
RTD
   
\end{abstract}

\section{Introduction}
\label{sec:intro}

In the last two decades, the Wigner function approach~\cite{Wigner1932,WeinbubFerry2018} has provided a powerful tool for studying quantum effect in various electronic devices, such as the resonant tunneling diodes (RTDs)~\cite{VandeputSoreeMagnus2017} and the metal oxide semiconductor filed effect transistors (MOSFETs)~\cite{GehringKosina2005}. A coupled system of the Wigner equation and the Poisson equation is usually adopted 
for taking the space charge effects into account. Finite difference methods were often used to obtain numerical solutions of the Wigner equation~\cite{Frensley1987,th:Biegel1997} as well as of the Wigner-Poisson (WP) system~\cite{th:Zhao2000,JiangCaiTsu2011}, and several spectral methods were also tried~\cite{Ringhofer1992, ArnoldRinghofer1996,VandeputSoreeMagnus2017}. 
In order to accurately capture 2-D quantum transport in a double-gate MOSFET (dgMOSFET), 
the WP system in 4-D phase space is required to be integrated with high resolution. 
However, all above-mentioned numerical methods were implemented in 2-D phase space,
and highly accurate deterministic numerical methods for the WP system in 4-D phase space are very few up to now.
This paper is intended to fill this gap by exploiting a recently developed 
operator splitting spectral method for the 4-D Wigner equation in quantum double-slit interference~\cite{ChenShaoCai2019}. 
Specifically, we will take advantage of the operator splitting spectral method to solve the 4-D  Wigner equation, 
in which the semi-discrete models resulted from spectral expansion in phase space for the sub-equations 
have analytical solutions, and continue to use a Chebyshev spectral method to solve the 2-D Poisson equation. 

Detailed benchmark tests are performed with the Gaussian barrier scattering in 2-D and 4-D phase space, and demonstrate that the proposed operator splitting spectral method indeed has a spectral accuracy in phase space and a fourth-order accuracy in time. We also show that the electric field induced by the space charge has a great effect on the rate of quantum tunneling.  
After calibration, we apply our high-resolution solver into simulating RTD and dgMOSFET.
Numerical experiments show that the steady states can be well developed from long-time simulations and the corresponding current-voltage (I-V) curves are numerically converged as the number of collocation points increases. 

The remainder of this paper is organized as follows. Section~\ref{sec:WPE} briefs the WP system. Section~\ref{sec:num-sche} presents the operator splitting spectral method. Section~\ref{sec:result} conducts benchmark tests with the Gaussian barrier scattering. Simulations and discussions of RTD and dgMOSFET are given in Sections~\ref{sec:rtd} and  \ref{sec:mosfet}, respectively. The paper is concluded in Section~\ref{sec:conclude} with a few remarks.

\section{The Wigner-Poisson system}
\label{sec:WPE}

The Wigner function $f(\bm x, \bm k, t)$ living in $2d$-D phase space: $(\bm x,\bm k) \in \mathbb{R}^{2d}$ with position $\bm x\in\mathbb{R}^d$ and wavevector $\bm k\in\mathbb{R}^d$,  obeys the following Wigner equation \cite{Wigner1932} 
\begin{equation}
  \label{eq:wigner}
  \frac{\partial}{\partial t}f(\bm x,\bm k,t)+\frac{\hbar \bm k}{m}\cdot\nabla_{\bm x}f(\bm x,\bm k,t) = \Theta_V[f](\bm x,\bm k,t),
\end{equation}
where $d$ gives the dimension of position space, $t$ denotes the time, $\hbar$ is the reduced Planck constant, $m$ is the mass, 
and  
$\Theta_V[f]$ is the so-called nonlocal pseudo-differential operator containing all the quantum information: 
\begin{align}
  \Theta_V[f](\bm x,\bm k,t) &= \int_{\mathbb{R}^d}\D \bm k'f(\bm x,\bm k',t)V_w(\bm x,\bm k-\bm k',t),\\
  V_w(\bm x,\bm k,t) &=\frac{1}{\mi\hbar (2\pi)^d}\int_{\mathbb{R}^d}\D \bm y \me^{-\mi \bm k\bm y}\left [ V(\bm x+\frac{\bm y}{2},t)-V(\bm x-\frac{\bm y}{2},t)\right ].
\end{align}
Here $V(\bm x,t)$ gives the external potential, and can be rewritten into $V(\bm x,t) = V_b(\bm x) + V_e(\bm x,t)$
when taking the space charge effects into account,   
where  $V_b(\bm x)$ denotes the conduction band potential and $V_e(\bm x,t)$ the effective electric potential.
Actually, $V_e(\bm x,t)$ can be determined by a Poisson equation with the electron density as its source term:  
\begin{equation}\label{eq:poisson}
  -\nabla_{\bm x}(\epsilon(\bm x)\nabla_{\bm x}) V_e(\bm x,t) = -q_e[n(\bm x,t) - N_d(\bm x)],
\end{equation}
where $q_e$ denotes the positive electron charge, $\epsilon(\bm x)$ is the dielectric constant, $N_d(\bm x)$ is the doping density and $n(\bm x,t)$ denotes the density of electrons given by
\begin{equation}
  \label{eq:density_n}
  n(\bm x,t) = \int_{\mathbb{R}^d} \D \bm k f(\bm x,\bm k, t).
\end{equation}
And, the current density ${\bm J}(\bm x,t)$ can be further calculated by 
\begin{equation}
  \label{eq:current_j}
  {\bm J}(\bm x,t) =\int_{\mathbb{R}^d} \D \bm k  \frac{\hbar\bm k}{m}f(\bm x,\bm k,t).
\end{equation}

\begin{figure}
  \centering
  \includegraphics[width=0.65\textwidth,height = 0.38\textwidth]{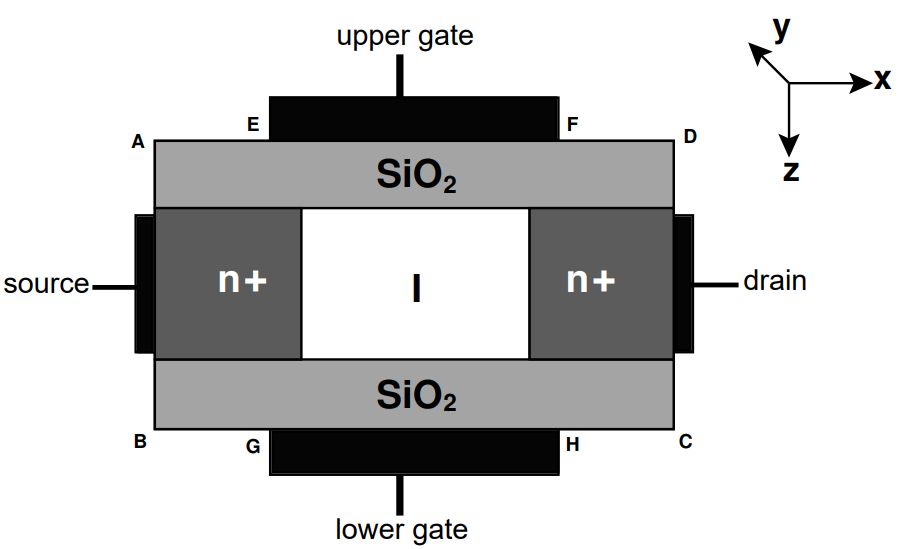}
  \caption{\small A 16 nm dgMOSFET structure \cite{RenVenugopalGoasguenDattaLundstrom2003}. The gate length $L_G$, equivalent gate oxide thickness $EOT$ and silicon channel thickness $T_{si}$ are 6 nm, 1 nm and 3 nm, respectively. The source and drain doping is $5\times 10^{19}~\text{cm}^{-3}$. The transistor is assumed to be wide, i.e., the $y$-direction is treated as infinite long.}
  \label{fig:mosfet_device}
\end{figure}

In this work, we focus on developing a high-resolution solver for the WP system in 4-D phase space (i.e., $d=2$) and let  $\bm x = (x,z)$, $\bm k = (k_x,k_z)$. In particular, our target is to simulate the dgMOSFET (as shown in Fig.\ref{fig:mosfet_device}) and the working-equations read 
\begin{equation}\label{eq:wigner_poisson_2d}
  \begin{split}
    (a)~~~& \frac{\partial }{\partial t}f(\bm x,\bm k,t) + \frac{\hbar \bm k}{m}\cdot\nabla_{\bm x}f(\bm x,\bm k,t) = \Theta_V[f](\bm x,\bm k,t),\\
    (b)~~~& f(\bm x,\bm k,t=0) = f_0(\bm x,\bm k), \\
    (c)~~~& f(x_l,z,\bm k,t) = f_{lx}(z, \bm k,t),\quad k_x>0, \quad f(x_r,z,\bm k,t) = f_{rx}(z,\bm k,t),\quad k_x<0,\\
    (d)~~~& f(x,z_l,\bm k,t) = 0,\quad k_z>0, \quad f(x,z_r,\bm k,t) = 0, \quad k_z<0,\\
    (e)~~~& -\Delta V_e(x,z,t) = q_e(-n(x,z,t) + N^+_D)/\epsilon,\\
    (f)~~~& V_e(x,z,t) = V_{gl},  ~~  \text{at GH}, \quad V_e(x,z,t)=V_{gu}, ~~ \text{at EF},\\
       & \partial_z V_e(x,z,t) = 0,\quad  \text{at BG, HC, AE, FD},\\
    (g)~~~&V_e(x,z,t) = 0, ~~\text{at AB},  \quad V_e(x,z,t) = -V_{ds},~~\text{at CD},
  \end{split}
\end{equation}
where we have chosen the commonly used inflow boundary conditions for the Wigner equation \cite{Frensley1987, ShaoLuCai2011},  
and mixed boundary conditions for the Poisson equation: the Dirichlet boundary at the source/drain in $x$-direction and the gates in $z$-direction plus the Neumann boundary at the Oxide/Air interfaces in $z$-direction, the computational domain for position is $\Omega_{\bm x}=[x_l,x_r]\times [z_l,z_r]$,  $V_{gu}$ and $V_{gl}$ give the upper and lower gate voltage, respectively, $V_{ds}$ is the source/drain bias potential, and $N_D^+$ refers to the ionized donor doping concentration.

\section{Numerical methods }
\label{sec:num-sche}

Considering the decay property of the Wigner function when $|\bm k|\rightarrow +\infty$, a simple nullification outside a sufficiently large $\bm k$-domain $\Omega_{\bm k}$ is usually adopted \cite{XiongChenShao2016, ChenXiongShao2019, ChenShaoCai2019}, thus we are in fact using a truncated pseudo-differential operator $\Theta^T_V[f]$ in $\bm k$-space as follows
\begin{equation}\label{eq:truncated_pdt}
  \begin{split}
    \Theta^T_V[f](x,z,k_x,k_z,t) &=  \iint_{\Omega_{\bm k}}\D k_x'\D k_z'f(x,z,k_x',k_z',t)\tilde{V}_w(x,z,k_x-k_x',k_z-k_z',t), \\
    \tilde{V}_w(x,z,k_x,k_z,t) &= \frac{\Delta y_x\Delta y_z}{\mi\hbar(2\pi)^2} \sum_{\mu=-\infty}^{+\infty}\sum_{\nu=-\infty}^{+\infty}D_V(x,z,y_{x,\mu},y_{z,\nu},t) \me^{-\mi k_x y_{x,\mu} - \mi k_z y_{z,\nu}},\\
    D_V(x,z,y_x,y_z,t) & = V(x+\frac{y_x}{2},z+\frac{y_z}{2}, t) - V(x-\frac{y_x}{2},z-\frac{y_z}{2},t),
  \end{split}
\end{equation}
where $\Omega_{\bm k} = [k_{x,\min}, k_{x,\max}]\times [k_{z,\min}, k_{z,\max}]$ and $y_{x,\mu} = \mu\Delta y_x$, $y_{z,\nu} = \nu \Delta y_z$ with $\Delta y_x$, $\Delta y_z$ being the spacing, which satisfy $ \Delta y_i = 2\pi/L_{k_i}  $ with $L_{k_i}=k_{i,\max}- k_{i,\min}$  for $i=x, z$ in this paper.

\subsection{Solving the 4-D Wigner equation}

The operator splitting spectral method developed in \cite{ChenShaoCai2019} for simulating the quantum double-slit interference is employed here for solving the 4-D Wigner equation. A brief description is given below and the interested readers are referred to \cite{ChenShaoCai2019} for more details.

An $s$-stage exponential operator splitting method for the Wigner equation given in Eq.~\eqref{eq:wigner} reads
\begin{equation}\label{OS}
  f^{n+1}(\bm x, \bm k) = \me^{\Delta t(A+B)}f^n(\bm x,\bm k) = \prod_{j=1}^s\me^{a_j\Delta tA}\me^{b_j\Delta t B}f^{n}(\bm x,\bm k) + \mathcal{O}(\Delta t^{s+1}),
\end{equation}
where $f^{n}(\bm x,\bm k):= f(\bm x,\bm k,t^n)$ denotes the exact solution at time $t^n:=n\Delta t$ and $\prod_{j=1}^s\me^{a_j\Delta tA}\me^{b_j\Delta t B}f^{n}(\bm x,\bm k)$ gives the corresponding numerical solution. Here $A$, $B$ are the convection operator and pseudo-differential operator, which correspond to two sub-equations of the Wigner equation, respectively:
\begin{equation}\label{eq:wigner_split}
\left\{
\begin{split}
\text{(A)}\quad \frac{\partial}{\partial t}f(\bm x,\bm k,t)  & = -\frac{\hbar \bm k}{m}\cdot\nabla_{\bm x}f(\bm x,\bm k,t),  \\
\text{(B)}\quad \frac{\partial}{\partial t}f(\bm x,\bm k,t)  & =  \Theta^T_V[f](\bm x,\bm k,t).
\end{split}
\right.
\end{equation}
We adopt the advective approach to march the sub-equation (A) in Eq.~\eqref{eq:wigner_split} strictly along the characteristic lines as follows
\begin{equation}
  \label{eq:split_A}
  f^{n+1}(\bm x,\bm k) = \me^{\Delta t A}f^n(\bm x,\bm k) = f^n(\bm x- \bm v \Delta t, \bm k), \quad \bm v = \frac{\hbar \bm k}{m},
\end{equation}
and the Chebyshev expansion of the Wigner function with respect to $\bm x$ is used to obtain function values at shifted points. 

Motivated by the intrinsic nature of Fourier transformation contained in the pseudo-differential term Eq.~\eqref{eq:truncated_pdt}, we use a Fourier spectral method to solve the sub-equation (B) in Eq.~\eqref{eq:wigner_split}.
 The interpolation operator $\mathcal{I}_{\bm{k, N}}$ reads
\begin{equation}
  \mathcal{I}_{\bm{k, N}}f(\bm x,\bm k,t) = \sum_{\nu_x=-N_x/2+1}^{N_x/2+1}\sum_{\nu_z=-N_z/2+1}^{N_z/2}a_{\nu_x,\nu_z}(\bm x,t)\psi_{\nu_x}(k_x)\psi_{\nu_z}(k_z). 
\end{equation}
where $\psi_{\nu_i}(k_i)  = \me^{2\pi\mi \nu_i(k_i-k_{i,\min})/L_{k_i}}$ are the Fourier basis functions and  $N_i$ the number of collocation points in $ k_i$-space for $i=x,z$. Substituting the interpolation function $\mathcal{I}_{\bm{k, N}}f(\bm x,\bm k,t)$ into the pseudo-differential term Eq.~\eqref{eq:truncated_pdt} also yields spectral approximation
\begin{align*}
 \Theta^T_V[f](\bm x,\bm k,t)& \approx \sum_{\nu_x=-N_x/2+1}^{N_x/2+1}\sum_{\nu_z=-N_z/2+1}^{N_z/2}c_{\nu_x,\nu_z}(\bm x)a_{\nu_x,\nu_z}(\bm x,t)\psi_{\nu_x}(k_x)\psi_{\nu_z}(k_z),\\
 c_{\nu_x,\nu_z}(x,z) &= \frac{1}{\hbar}D_V(x,z,y_{x,\nu_x},y_{z,\nu_z}).
\end{align*}
Accordingly, the orthogonal relation of the Fourier basis functions implies
\[
  \frac{\partial}{\partial t} a_{\nu_x,\nu_z}(\bm x,t) = c_{\nu_x,\nu_z}(\bm x)a_{\nu_x,\nu_z}(\bm x,t),
\]
the solution of which has the following explicit form
\[
   a^{n+1}_{\nu_x,\nu_z}(\bm x) = \me^{c_{\nu_x,\nu_z}(\bm x)\Delta t}a^n_{\nu_x,\nu_z}(\bm x).
\]

To match with the spectral accuracy in phase space, we adopt a fourth-order splitting scheme with $s=4$ in Eq.~\eqref{OS}:
\begin{align*}
  & a_1 = a_4 = \frac{1}{2(2-\sqrt[3]{2})}, \quad a_2 = a_3 = \frac{1-\sqrt[3]{2}}{2(2-\sqrt[3]{2})},\\
  & b_1 = b_3 = \frac{1}{2-\sqrt[3]{2}}, \quad  b_2 = -\frac{\sqrt[3]{2}}{2-\sqrt[3]{2}}, \quad b_4 = 0. 
\end{align*}
In subsequent numerical experiments, we usually choose $N_x =N_z := N$ for convenience.

\subsection{Solving the 2-D Poisson equation}
\label{sec:2d}

The Chebyshev expansion in $\bm x$ direction continues to be used to solve the Poisson equation in $[x_l,x_r]\times [z_l,z_r]$:
\begin{equation}\label{eq:poisson_2d}
  \Delta V(x,z) = r(x,z)
\end{equation}
with Dirichlet boundary conditions:
\begin{equation}\label{eq:bound_2d}
  V(x,z_l) = V_{gl}(x),\quad V(x,z_r) = V_{gu}(x), \quad  V(x_{l},z) = 0, \quad  V(x_{l},z) = -V_{ds},
\end{equation}
where the function $V_{gl}(x)$ (resp. $V_{gu}(x)$) reduces to the lower (resp. upper) gate voltage $V_{gl}$ at the gate,  
and vanishes otherwise.  In order to achieve the spectral convergence, 
we use a cubic polynomial to smoothly connect, for example,  $V_{gl}$  to $0$.

We assume that $V(x,z)$, $r(x,z)$, $\Delta V(x,z)$, $V_{gl}(x)$, $V_{gu}(x)$ can be approximated by the truncated Chebyshev series as follows
\begin{align*}
  V(x,z) &\approx \sum_{n=0}^M\sum_{m=0}^Ma_{nm}\phi_{n}(x)\phi_{m}(z), \\
  r(x,z) &\approx \sum_{n=0}^M\sum_{m=0}^Mb_{nm}\phi_n(x)\phi_m(z),\\
  \Delta V(x,z) &\approx \sum_{n=0}^M\sum_{m=0}^M a^{(2)}_{nm}\phi_n(x)\phi_m(z),\\ 
   V_{gu}(x) & \approx \sum_{n=0}^M b^u_n \phi_n(x), \quad
    V_{gl}(x) \approx \sum_{n=0}^M b^l_n \phi_n(x),
\end{align*}
where $\phi_n$ give the Chebyshev polynomials of the first kind.

For simplicity, we suppose that the numbers of collocation points in $x$ and $z$ directions are even and the same, denotes by $M$. It can be readily verified that the expansion coefficients satisfy the following relationships
\begin{align*}
  a^{(2)}_{nm} &= \frac{1}{c_n}\sum_{p=n+2:2:M}p(p^2-n^2)a_{pm}+\frac{1}{c_m}\sum_{q=m+2:2:M}q(q^2-m^2)a_{nq}, 
\end{align*}
where $c_0 = 2$ and $c_n = 1$ for $n\ge 1$.

The collocation equations for $\{a_{nm}\}$ that follow from Eqs.~\eqref{eq:poisson_2d}-\eqref{eq:bound_2d} are then
\begin{equation}\label{eq:poi_2d}
  \frac{1}{c_n}\sum_{p=n+2:2:M}p(p^2-n^2)a_{pm}+\frac{1}{c_m}\sum_{q=m+2:2:M}q(q^2-m^2)a_{nq} = b_{nm}, \quad 0\leq n,m \leq M-2,
\end{equation}
\begin{equation}\label{eq:bound_x}
\begin{split}
 \sum_{n~\text{odd}}a_{nm} = 0,\quad \sum_{n~\text{even}}a_{nm} = 0,\quad m = 1,\ldots, M,
 \end{split}
\end{equation}
\begin{equation}\label{eq:bound_z}
    \sum_{m~ \text{odd}}a_{nm} = (b^u_n-b^l_n)/2,\quad \sum_{m~\text{even}}a_{nm} = (b^u_n+b^l_n)/2,\quad n = 0,1,\ldots, M.
  \end{equation}
We define the column vectors $X_i$, $B_i$ for $i = 0,1,\ldots,M$ by
\begin{align*}
  X_{i} &= (a_{0i},a_{1i},\cdots,a_{M-1,i},a_{Mi})^{\top},\quad i = 0,1,\ldots, M,\\
  B_{0} &= ((b^u_0+b^l_0)/2,(b^u_1+b^l_1)/2,\cdots,(b^u_M+b^l_M)/2)^{\top}, \\
  B_{1} &= ((b^u_0-b^l_0)/2,(b^u_1-b^l_1)/2,\cdots,(b^u_M-b^l_M)/2)^{\top}, \\
  B_{i} &= (0,0,b_{0i},b_{1i},\cdots,b_{M-3,i}, b_{M-2,i})^{\top},\quad i = 2,3,\ldots, M,
\end{align*}
and let $P$ be the $(M+1)\times (M+1)$ matrix as shown in Eq.~\eqref{eq:P} in Appendix.
Consequently, Eqs.~\eqref{eq:poi_2d}-\eqref{eq:bound_z} can be rewritten into

{\small
\begin{equation}\label{eq:matrix_mix}
  \left(
  \begin{array}{cccccccc}
    I & 0 & I & 0 & \cdots & I & 0 & I \\
    0 & I & 0 & I & \cdots & 0 & I & 0 \\
    P & 0 & A_{02}& 0 &\cdots & A_{0,M-2}& 0 & A_{0M}\\
    0 & P & 0 & A_{13} & \cdots & 0 & A_{1,M-1} & 0\\
    \vdots&\vdots&\vdots&\vdots&\vdots&\vdots&\vdots&\vdots\\
    0 & 0 & 0 & 0 & \cdots  & 0 & A_{M-3,M-1}  & 0\\
    0 & 0 & 0 & 0 & \cdots & P & 0 & A_{M-2,M} \\
  \end{array}\right)
  \left(
  \begin{array}{c}
    X_0\\
    X_1\\
    X_2\\
    X_3\\
    \vdots\\
    X_{M-1}\\
    X_M\\
  \end{array}\right) = \tilde{F},
\end{equation}}

\noindent where $\tilde{F} = (B_0,B_1,\cdots,B_{M-1},B_M)^{\top}$ and $A_{ij}=c_ij(j^2-i^2)\tilde{I}$ with $\tilde{I}$ being the $(M+1)\times (M+1)$ matrix in Eq.~\eqref{eq:I_tilde} in Appendix. 

The solution process and related specific solution form are detailed in Appendix.
Although the size of the coefficient matrix in Eq.~\eqref{eq:matrix_mix} is $(M+1)^2\times(M+1)^2$, only the calculation of sub-matrices of order $(M+1)\times (M+1)$ is involved rather than directly inverting the original matrix. Therefore, the proposed Chebyshev spectral method for the 2-D Poisson equation is not only highly accurate but also efficient. It should be pointed out that we select the Dirichlet boundary condition above just for an example and the proposed numerical solver for the Poisson equation is able to deal with all kinds of boundary conditions.

In summary, we evolve  the WP system~\eqref{eq:wigner_poisson_2d}  in 4-D phase space as follows
\begin{itemize}
\item[Step I.] Calculate the potential $V_e(\bm x, t=0)$ with initial density $n(\bm x,t=0)$ via the 2-D Poisson equation (Eq.~\eqref{eq:poisson}) by using Chebyshev spectral methods;
\item[Step II.] Using the obtained potential $V_e(\bm x,t=0)$ to solve the time-dependent 4-D Wigner equation with operator splitting spectral method to obtain $f(\bm x,\bm k, \Delta t)$ and then to calculate the density $n(\bm x,\Delta t)$ via Eq.~\eqref{eq:density_n};
\item[Step III.] Calculate the potential $V_e(\bm x, \Delta t)$ with the density $n(\bm x,\Delta t)$, repeat Step I and Step II until to the final time $t_f$.    
\end{itemize}

\section{Calibration}
\label{sec:result}
In this section, we first would like to verify the convergence rate and efficiency of the proposed solver. 
The $L^2$-error $\varepsilon_2(t)$ and $L^{\infty}$-error $\varepsilon_{\infty}(t)$:
\begin{align}
  \varepsilon_2(t) =&(\iint_{\Omega}(f^{\text{num}}(\bm x,\bm k,t)-f^{\text{ref}}(\bm x,\bm k,t))^2\D \bm x \D \bm k)^{1/2},\\
  \varepsilon_{\infty}(t) = & \max_{(\bm x,\bm k)\in\Omega}\{|f^{\text{num}}(\bm x,\bm k, t) - f^{\text{ref}}(\bm x,\bm k,t)|\},
\end{align}
are employed to study the convergence rate in terms of the number of collocation points and the time step, where $\Omega=\Omega_{\bm k}\times \Omega_{\bm x}$ gives the computational domain in 4-D phase space, $f^{\text{num}}$ and $f^{\text{ref}}$ denote the numerical solution and reference solution, respectively.
To conveniently visualize  the 4-D Wigner function, we plot the reduced 2-D Wigner function \cite{XiongChenShao2016} in this paper as follows
\begin{equation}
  \label{eq:reduce_W}
  F(q,k,t):= \iint_{\Omega_z\times\Omega_{k_z}}\D z\D k_zf(q,z,k,k_z,t) +  \iint_{\Omega_x\times\Omega_{k_x}}\D x\D k_xf(x,q,k_x,k,t).
\end{equation}
In addition, the Chebyshev collocation points in $\bm x$-direction for the 4-D Wigner equation and 2-D Poisson keep the same, which may avoid additional interpolations when calculating the electron density $n(\bm x, t)$ in Eq.~\eqref{eq:poisson}.

\vspace{0.5cm}

{\bf $\bullet$ A 2-D Poisson equation}

\vspace{0.5cm}

Consider the Poisson equation in $\Omega_{\bm x} = [-5,5]\times[-5,5]$ with mixed boundary conditions as follows
\begin{equation*}
  \begin{split}
   & \Delta V(x,z) = r(x,z), \\
   & \partial_{x}V(-5,z) = g_1(z),  \quad  \partial_x V(5,z) = g_2(z), \\
   & V(x,-5) = g_3(x),   \quad   V(x,5) = g_4(x),
  \end{split}
\end{equation*}
the reference solution of which is $V(x,z) = (x^2+z^2)\me^{x^2+z^2}$ and the right terms are
\begin{align*}
  &r(x,z) = [4-12(x^2+z^2)+4(x^2+z^2)^2]\me^{-(x^2+z^2)},\\
  &g_1(z)=-g_2(z) = 10(z^2+25)\me^{-25-z^2},\\
  &g_3(x)=g_4(x)=(25+x^2)\me^{-25-x^2}.
\end{align*}

Table \ref{tab:poisson_test} gives the calculation time (the second column) and $L^\infty$-error (the third column) under the different number of collocation points (the first column). We set $M = 8$, 16, 32, 64 and $128$. When $M$ is equal to 64 and 128, the $L^\infty$-error has reached $10^{-14}$, but the calculation time only takes less than $0.1$ second. The calculation time in the Table \ref{tab:poisson_test} is the serial time with 1 CPU (Intel$^\circledR$ Core$^{\text{TM}}$ i7-8550U CPU @ 1.80GHz). That is, the proposed Chebyshev spectral method for the 2-D Poisson equation is not only highly accurate but also efficient. The right plot in Table~\ref{tab:poisson_test} clearly shows the spectral convergence with respect to $M$.    

\renewcommand\arraystretch{1.5}
\begin{table}[ht!]
  \centering
  \caption{\small A 2-D Poisson equation: The calculation time (the second column) and $L^\infty$-error (the third column) under the different number of collocation points, i.e., $M$ (the first column). The right figure plots the spectral convergence with respect to $M$. }
  \begin{tabular}{cccc}
    \hline
    \hline
      $M$  &      time (s)  &  $L^\infty$-error  & \multirow{5}{*}{ 
       \includegraphics[width=0.4\textwidth,height=0.3\textwidth]{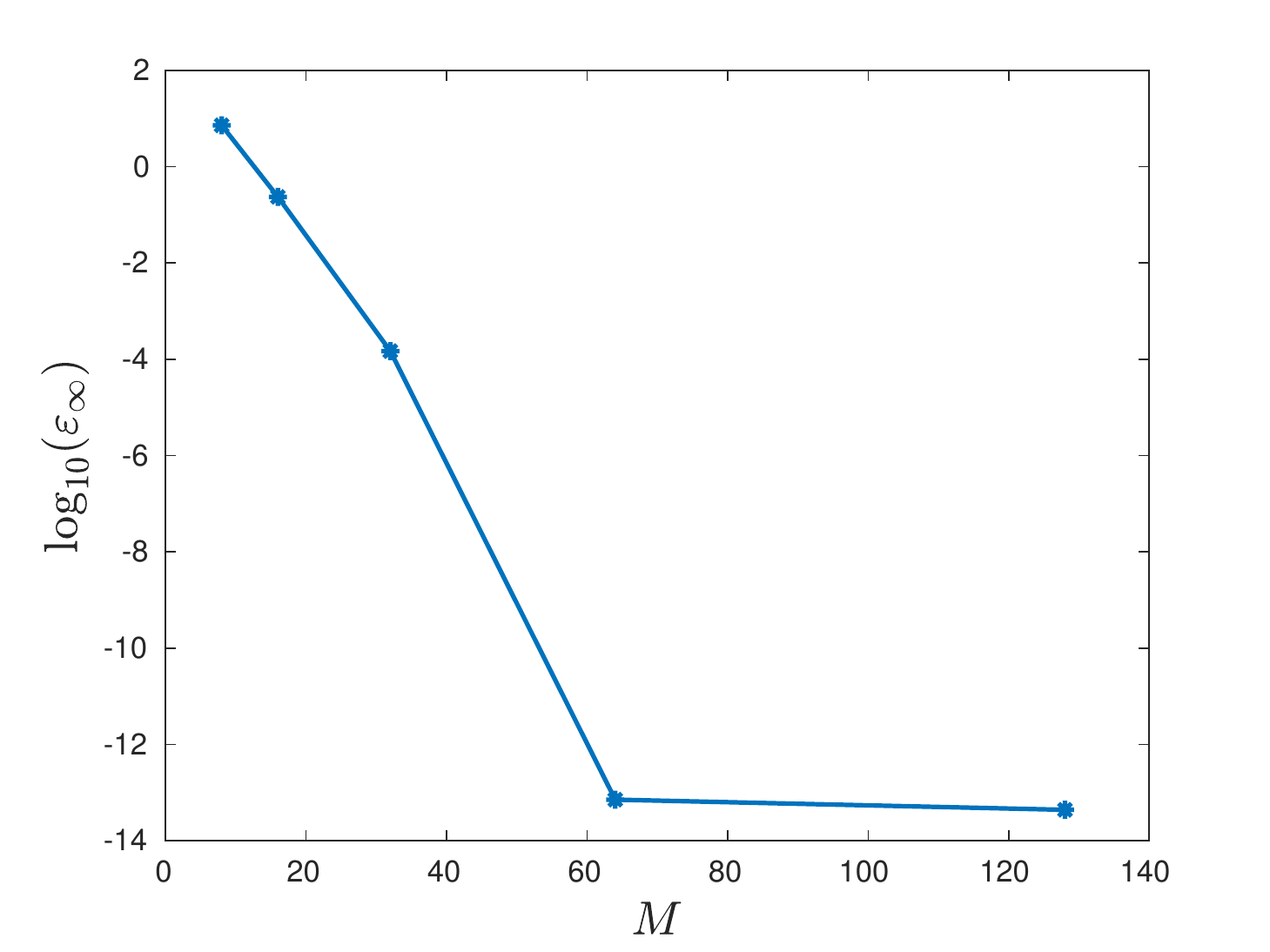}}  \\
    \cline{1-3}
        8  &  $4.4800\times 10^{-4}$  &   $7.1933$  & \\
    \cline{1-3}
       16  &  $1.1760\times 10^{-3}$  &   $0.2344$ &  \\
    \cline{1-3}
       32  &  $2.5050\times 10^{-3}$  &  $1.4719\times 10^{-4}$ & \\
    \cline{1-3}
       64  &  $1.1803\times 10^{-2}$  &  $7.1304\times 10^{-14}$ &\\
    \cline{1-3}
       128 &  $9.2475\times 10^{-2}$  &  $4.3883\times 10^{-14}$ &\\
    \hline
    \hline
  \end{tabular}
\label{tab:poisson_test}
\end{table}

\vspace{0.5cm}

{\bf $\bullet$ The WP system in 2-D phase space}

\vspace{0.5cm}

To further validate the overall performance of the operator splitting spectral method for the WP system, we simulate the Gaussian barrier scattering of the Gaussian wave packet (GWP) \cite{XiongChenShao2016,ChenShaoCai2019} to investigate its convergence rate. 
We first make tests in 2-D phase space: $\Omega_{k} = [-2\pi~\text{nm}^{-1}, 2\pi~\text{nm}^{-1}]$, $\Omega_{x} = [-25~\text{nm}, 25~\text{nm}]$, and adopt the initial GWP as
\begin{equation}\label{eq:GWP_1d}
  f_{0}(x,k) = \frac{1}{\pi}\exp[-\frac{(x-x_0)^2}{2a^2} - 2a^2(k-k_0)^2],
\end{equation}
where $x_0$ is the center, $a$ the minimum position spread and $k_0$ the initial wavenumber. The Gaussian barrier reads
\begin{equation}\label{eq:GB_1d}
  V_b(x) = H\exp[-\frac{(x-x_b)^2}{2\omega^2}]
\end{equation}
with $\omega = 1$ nm, $x_b = 0$ and $H = 2.3$ eV. The other parameters are: $x_0 = -10$ nm, $k_0 = 1.4~ \text{nm}^{-1}$, $a = \sqrt{2}$ nm, $\hbar = 1$ eV $\cdot$ fs, the effective mass $m_e = 1~\text{eV}\cdot\text{fs}^2\cdot\text{nm}^{-2}$ and the final time $t_f = 20$ fs.
The Poisson equation satisfies the Dirichlet boundary condition with bias potential $V_0 = 0.5$ eV, the dielectric constant $\epsilon = 10~\text{Fm}^{-1}$ and the doping density $N_d(x) = 0$.
                            
\begin{figure}[ht!]
  \includegraphics[width=0.32\textwidth,height=0.28\textwidth]{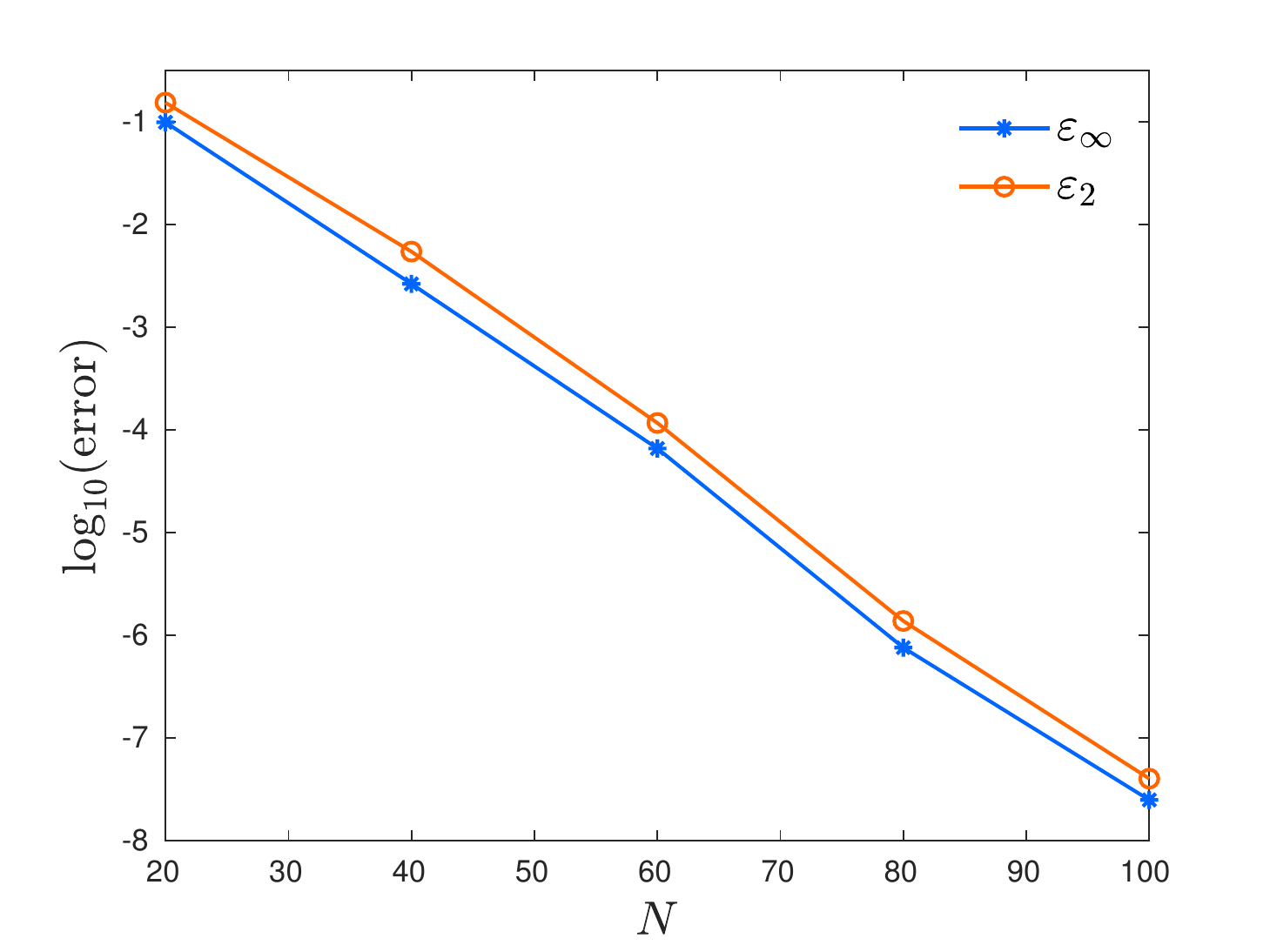}
  \includegraphics[width=0.32\textwidth,height=0.28\textwidth]{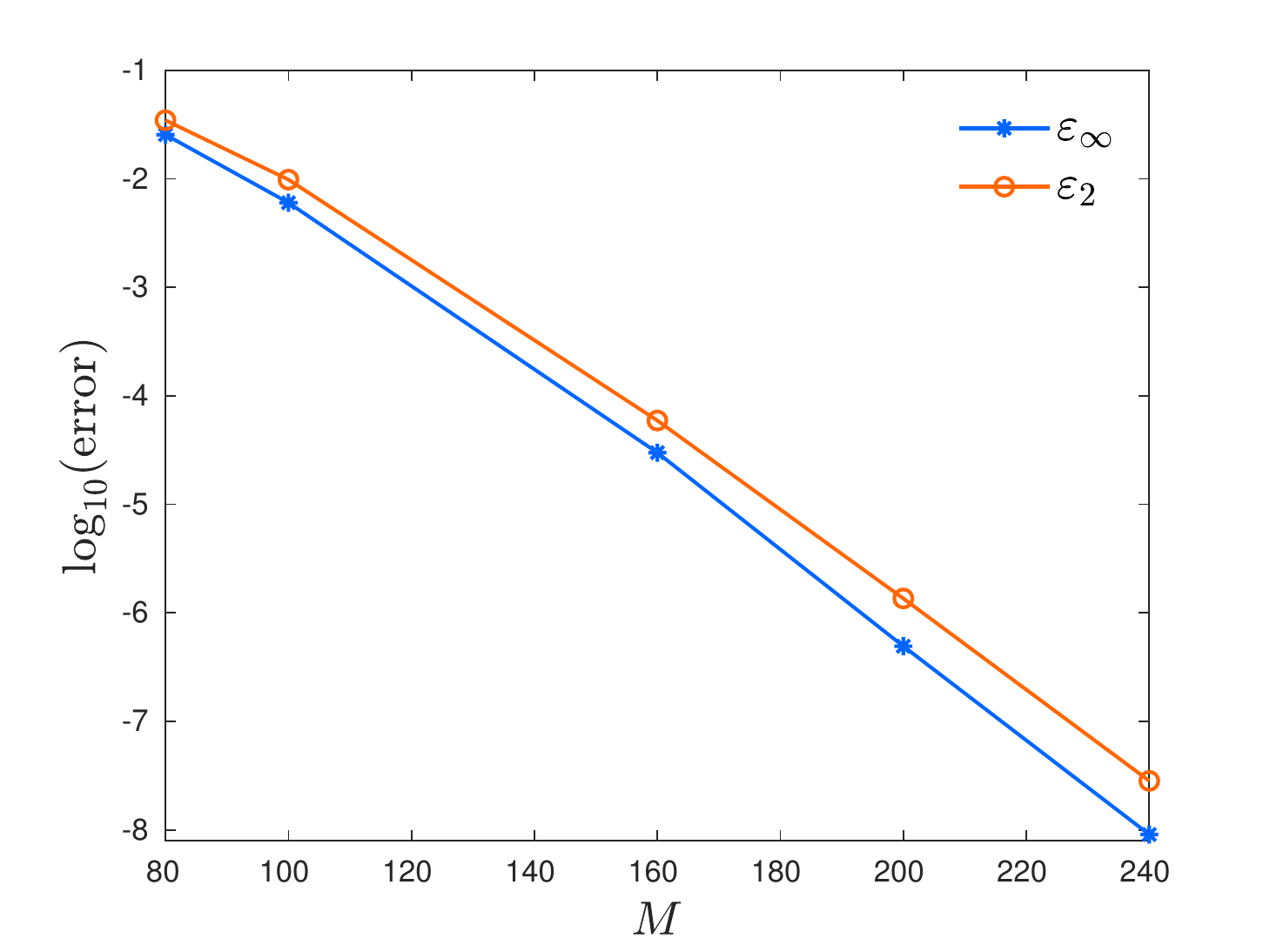}
  \includegraphics[width=0.32\textwidth,height=0.28\textwidth]{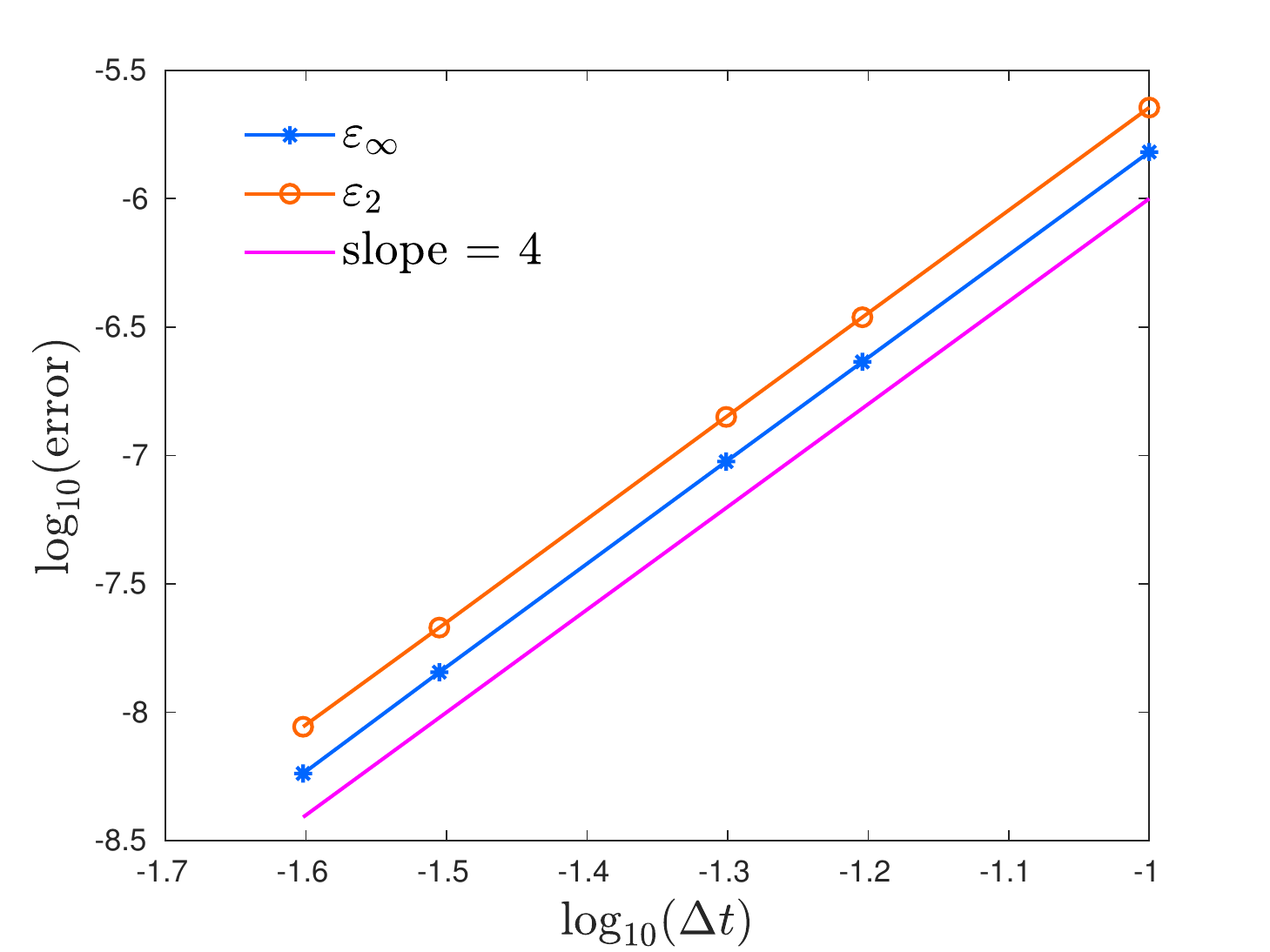}
  \caption{\small The WP system in 2-D phase space: Convergence rate with respect to $N$ (left) and $M$ (middle) and the time step $\Delta t$ (right). The spectral convergence in both $k$-space and $x$-space, and  the fourth-order accuracy against the time step are evident.}
  \label{fig:GBS_1d}
\end{figure}

In order to study the convergence rate with respect to $N$ (resp. M), we fix $M = 400$ (resp. $N = 128$) and $\Delta t = 0.01$ fs. As shown in the left and middle plots of Fig.~\ref{fig:GBS_1d}, the proposed splitting spectral method shows the spectral convergence with respect to both $N$ and $M$.  The right plot of Fig.~\ref{fig:GBS_1d} further displays the fourth-order convergence rate with respect to $\Delta t$ on a fixed mesh $(N, M) = (128,400)$.    

\begin{figure}[ht!]
  \includegraphics[width=0.32\textwidth,height=0.27\textwidth]{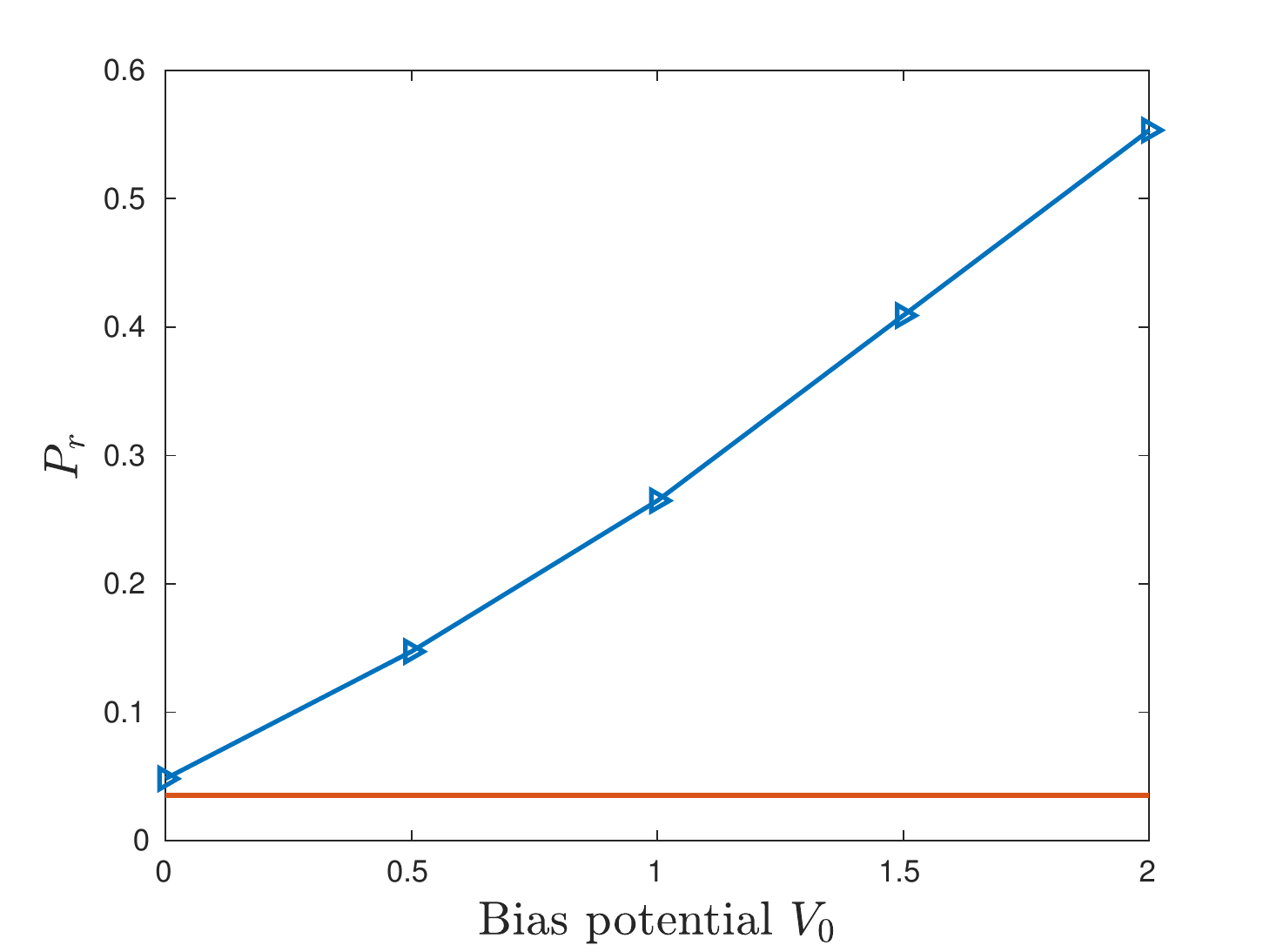}
  \includegraphics[width=0.32\textwidth,height=0.27\textwidth]{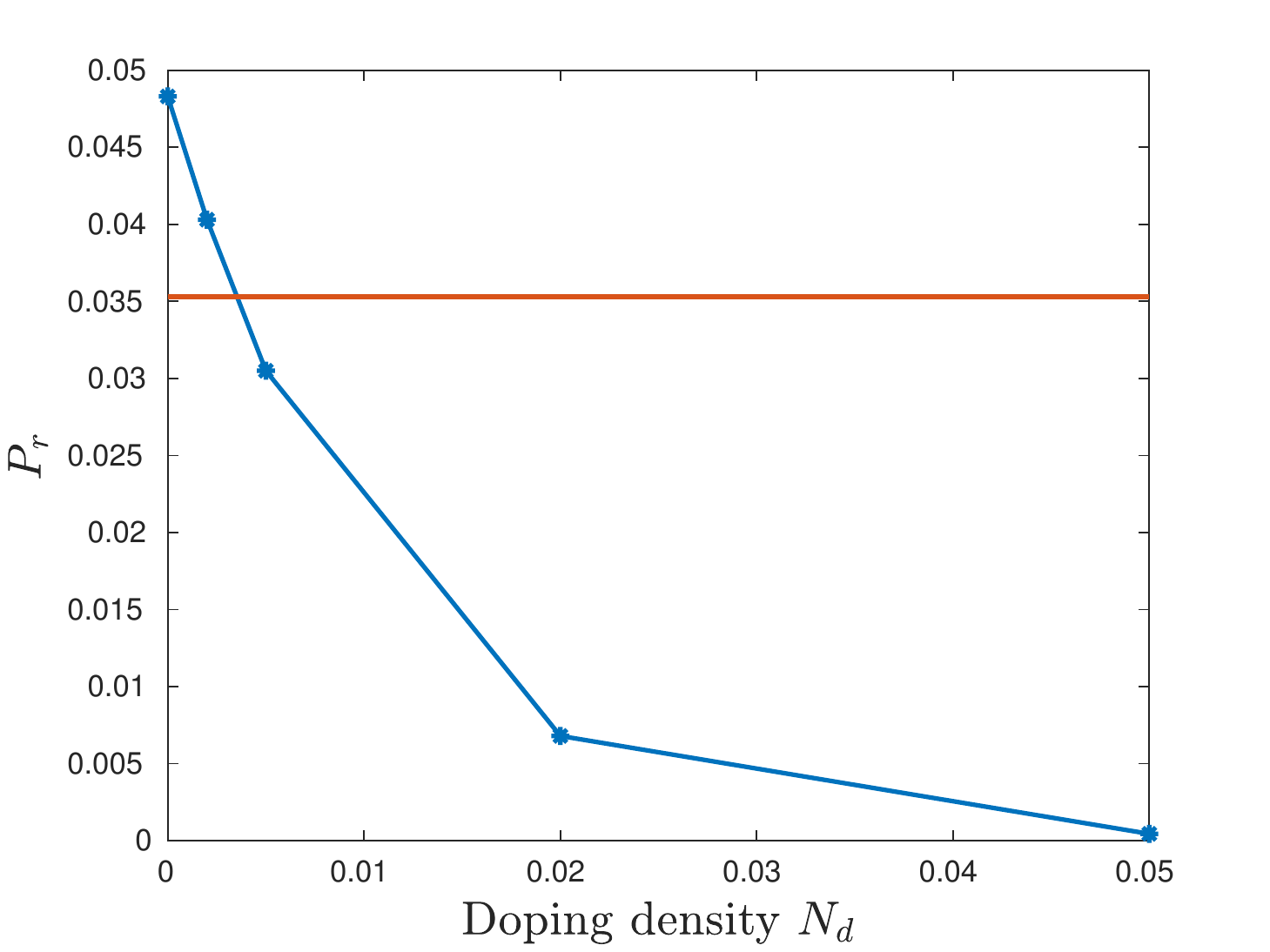}
  \includegraphics[width=0.32\textwidth,height=0.27\textwidth]{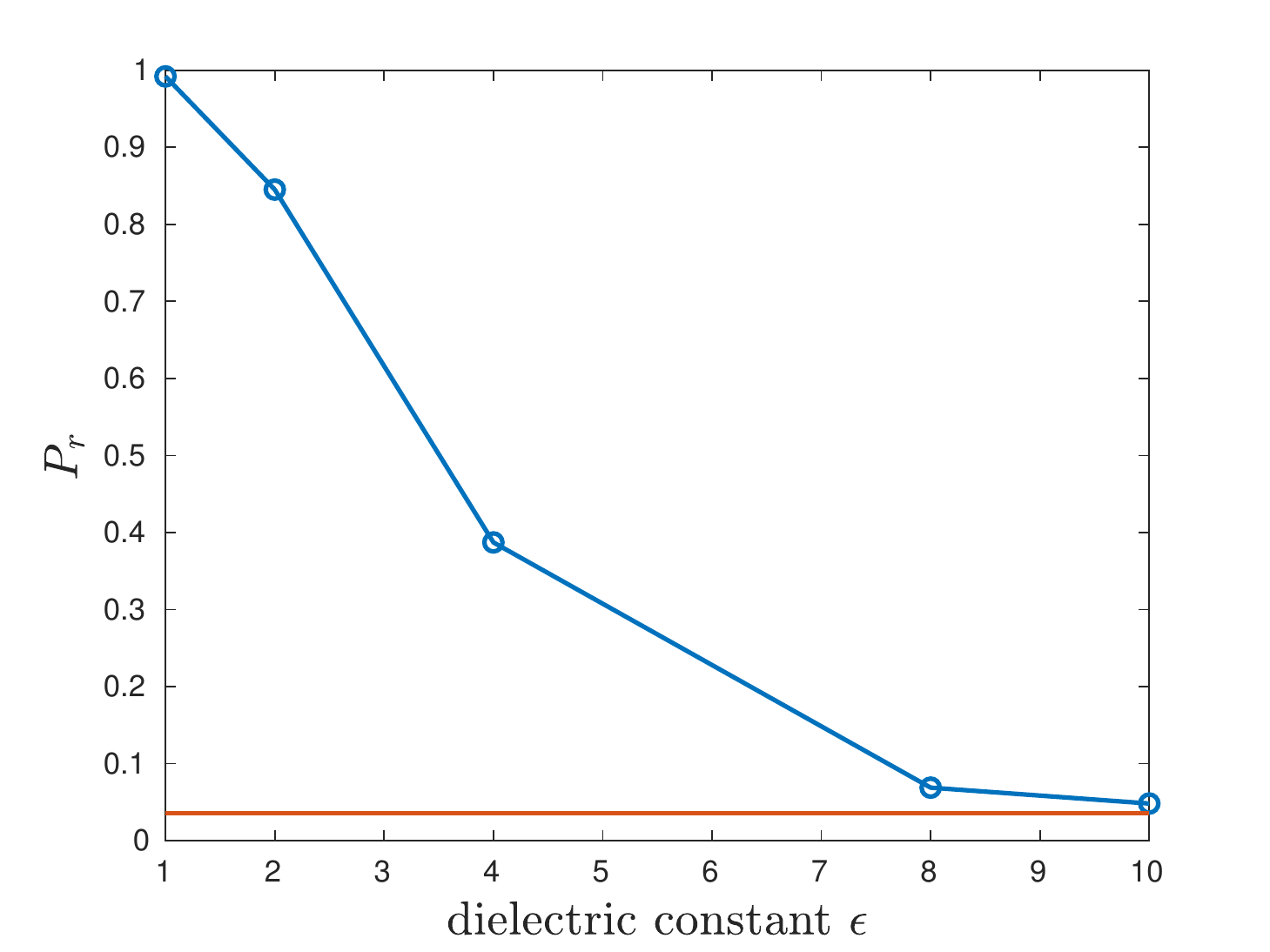}
  \caption{\small The WP system in 2-D phase space: Tunneling rate with/without coupling the Poisson equation. 
  The blue lines indicate the change of the tunneling rate with the bias potential $V_0$ (left), the doping density $N_d$ (middle) and the dielectric constant $\epsilon$ (right) when the Poisson equation is coupled. The red line represents the tunneling rate without coupling the Poisson equation.  }
  \label{fig:GBS_1d_pr}
\end{figure}

\begin{figure}[ht!]
   \subfigure[$t=8$, only Wigner equation.]{\includegraphics[width=0.48\textwidth,height=0.33\textwidth]{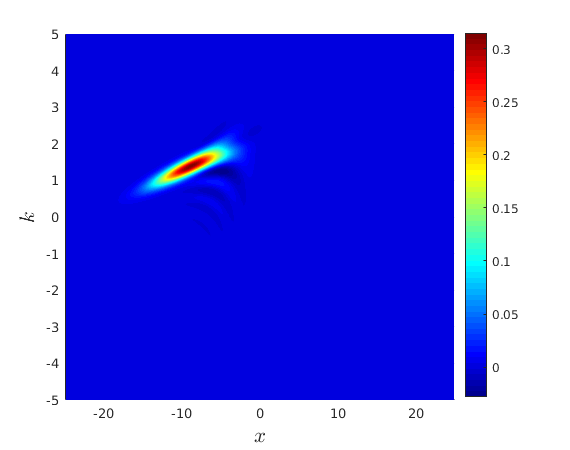}}
  \subfigure[$t=8$, WP system.]{ \includegraphics[width=0.48\textwidth,height=0.33\textwidth]{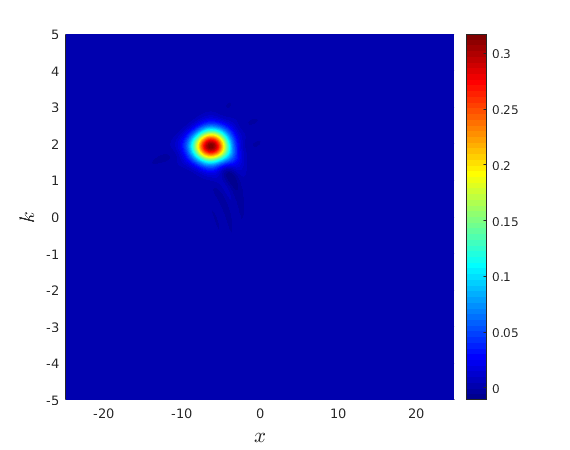}}
  
   \subfigure[$t=13$, only Wigner equation.]{\includegraphics[width=0.48\textwidth,height=0.33\textwidth]{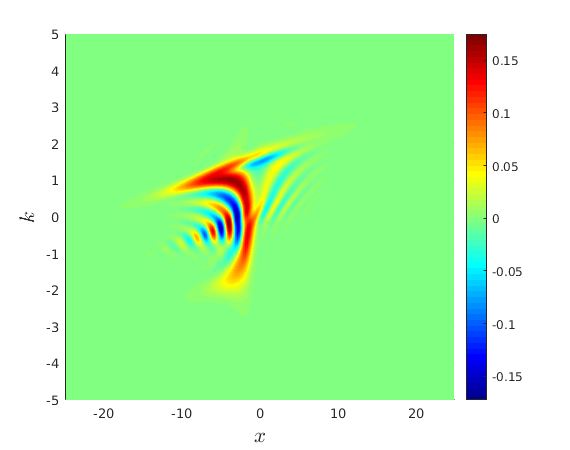}}
   \subfigure[$t=13$, WP system.]{\includegraphics[width=0.48\textwidth,height=0.33\textwidth]{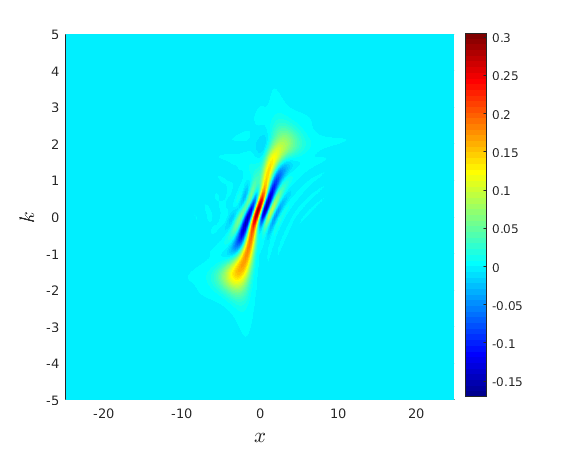}}
    
   \subfigure[$t=18$, only Wigner equation.]{ \includegraphics[width=0.48\textwidth,height=0.33\textwidth]{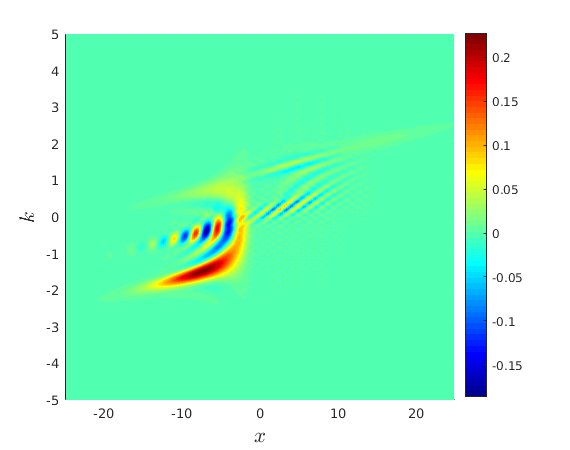}}
    \subfigure[$t=18$, WP system.]{\includegraphics[width=0.48\textwidth,height=0.33\textwidth]{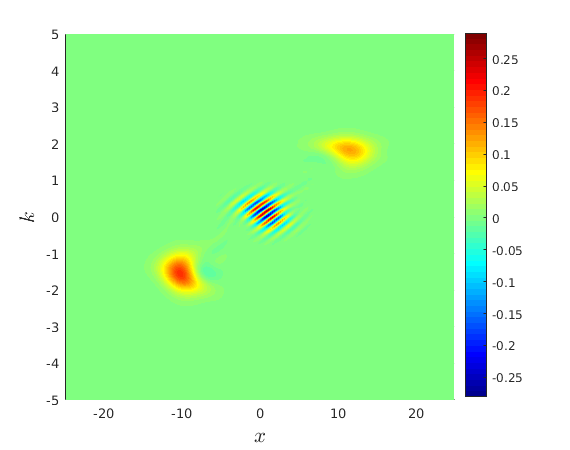}}
  \caption{\small The WP system in 2-D phase space: Wigner functions obtained by solving only the Wigner equation (left) and the WP system (right).
  We set $V_0 =0$, $N_d=0$, $\epsilon =4~\text{Fm}^{-1}$. It is clearly shown that the space charge effects helps GWP with its tunneling through the barrier.}
  \label{fig:GBS_1d_wigner}
\end{figure}

Next, we would like to use such Gaussian barrier scattering to study the effect of the space charge on quantum tunneling.  The tunneling rate $P_r(t)$ \cite{ChenXiongShao2019}
\[
    P_r(t) = \iint_{[0,x_r]\times \Omega_k}f(x,k,t) \D k\D x
\]
 is used to represent the part of GWP passing through the barrier. The mesh is fixed as $(N,M,\Delta t) = (128, 256, 0.025)$ and other parameters keep unchanged. Fixed $N_d = 0$ and $\epsilon =10~\text{Fm}^{-1}$, the tunneling rate is almost proportional to the bias potential $V_0 $ and higher than the value 0.0353 indicated by the red line, which is the rate for the case without coupling the Poisson equation, as shown in the left plot of Fig.~\ref{fig:GBS_1d_pr}. The middle plot of Fig.~\ref{fig:GBS_1d_pr}  gives the negative correlation between the tunneling rate and the doping density $N_d$ when the bias voltage $V_0 = 0$ and $\epsilon =10~\text{Fm}^{-1}$. 
The tunneling rate is much smaller than that for the case without coupling the Poisson equation when the doping density gets larger than a certain value (about $0.005$).  The right plot of Fig.~\ref{fig:GBS_1d_pr} also shows the negative correlation between the tunneling rate and the dielectric constant $\epsilon$, but the rate is always larger than that for the case without coupling the Poisson equation when fixed $V_0 = 0$, $N_d = 0$. 

We further compare the Wigner functions at instants $t= 8$, $13$, $18$ fs for only the Wigner equation (left) with those for the WP system (right) in Fig.~\ref{fig:GBS_1d_wigner} when setting $V_0 =0$, $N_d=0$, $\epsilon =4~\text{Fm}^{-1}$. 
We are able to clearly see there
that it is much easier for GWP to pass through the barrier when the Poisson equation accounting for the space charge effects is coupled.

\vspace{0.5cm}

{\bf $\bullet$ The WP system in 4-D phase space}

\vspace{0.5cm}

Now we will calibrate the proposed solver in 4-D phase space still with the Gaussian barrier scattering.
We choose the Gaussian barrier as
\begin{equation} \label{eq:barrier_2d}
  V_b(x,z) = 1.3 \exp(-x^2/2) + 1.3\exp(-z^2/2),
\end{equation}
and the initial GWP as
\begin{equation}\label{eq:initial_2d}
  f_0(x,z,k_x,k_z) = \frac{1}{\pi^2}\exp[-\frac{(x-x_0)^2}{2\sigma_x^2}-2\sigma_x^2(k_x-k_{x}^0)^2-\frac{(z-z_0)}{2\sigma_z^2} - 2\sigma_z^2(k_z-k_z^0)^2],
\end{equation}
where $x_0$, $z_0$ are the center, $k^0_{x/z}$ is the initial wavenumber and $\sigma_{x/z}$ is the minimum position spread. We set the parameters to be $\Omega_{\bm x} = [-20~\text{nm}, 20~\text{nm}]\times [-20~\text{nm}, 20~\text{nm}]$, $\Omega_k = [-2\pi~\text{nm}^{-1}, 2\pi~\text{nm}^{-1}]\times [-2\pi~\text{nm}^{-1}, 2\pi~\text{nm}^{-1}]$, $\sigma_x = \sigma_z = 1~\text{nm}$, $k_x^0=1.2~\text{nm}^{-1}$, $k_z^0=-1.2~\text{nm}^{-1}$, and $x_0=-6~\text{nm}$, $z_0=6~\text{nm}$. And we still choose $\hbar = 1$ eV $\cdot$ fs, $m = 1~\text{eV}\cdot\text{fs}^2\cdot\text{nm}^{-2}$, $\epsilon = 1~\text{Fm}^{-1}$, $N_d(x) = 0$ and $V_g$ = 0.5 V.

\begin{figure}[ht!]
  \includegraphics[width=0.48\textwidth,height=0.36\textwidth]{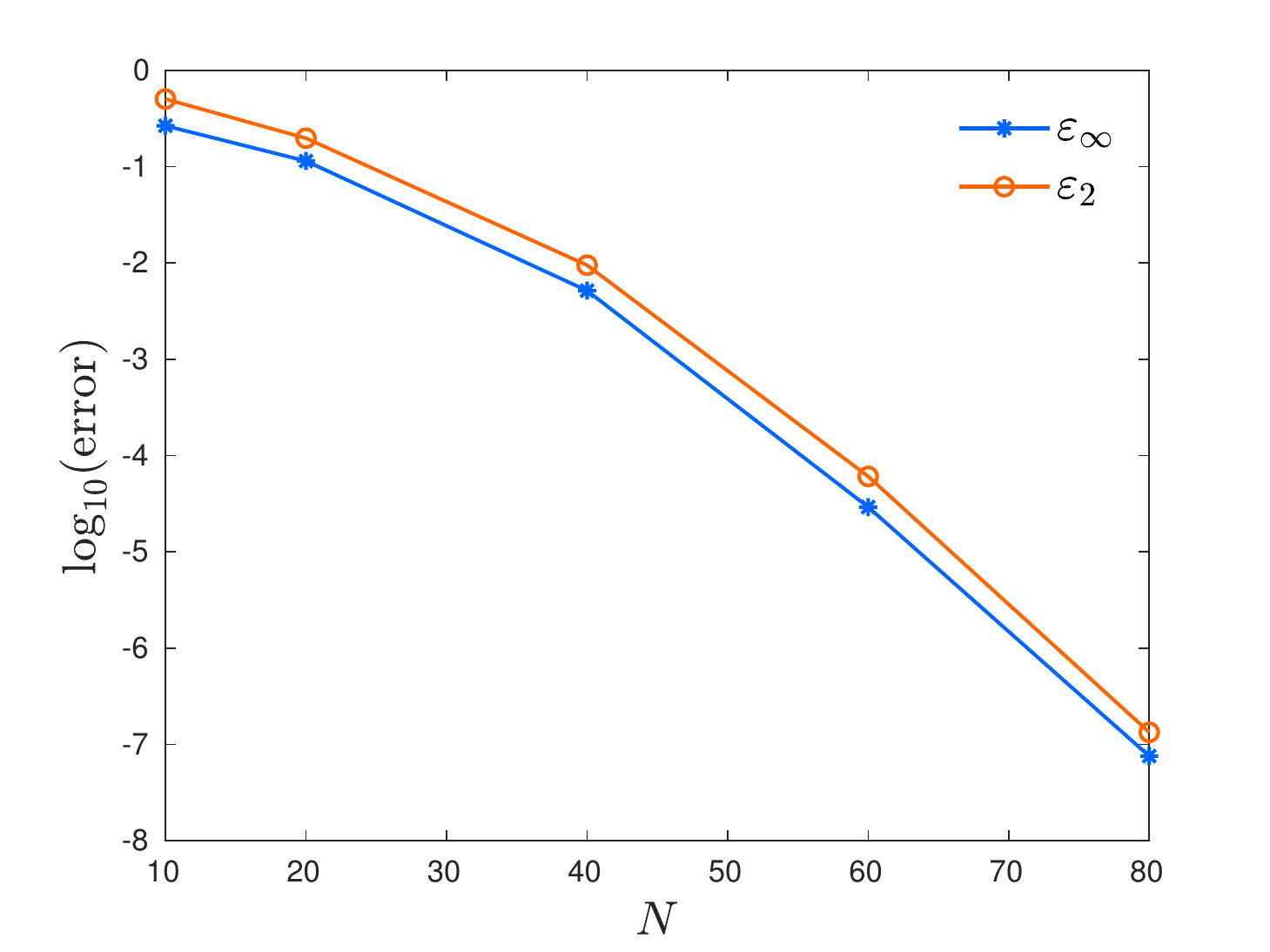}
  \includegraphics[width=0.48\textwidth,height=0.36\textwidth]{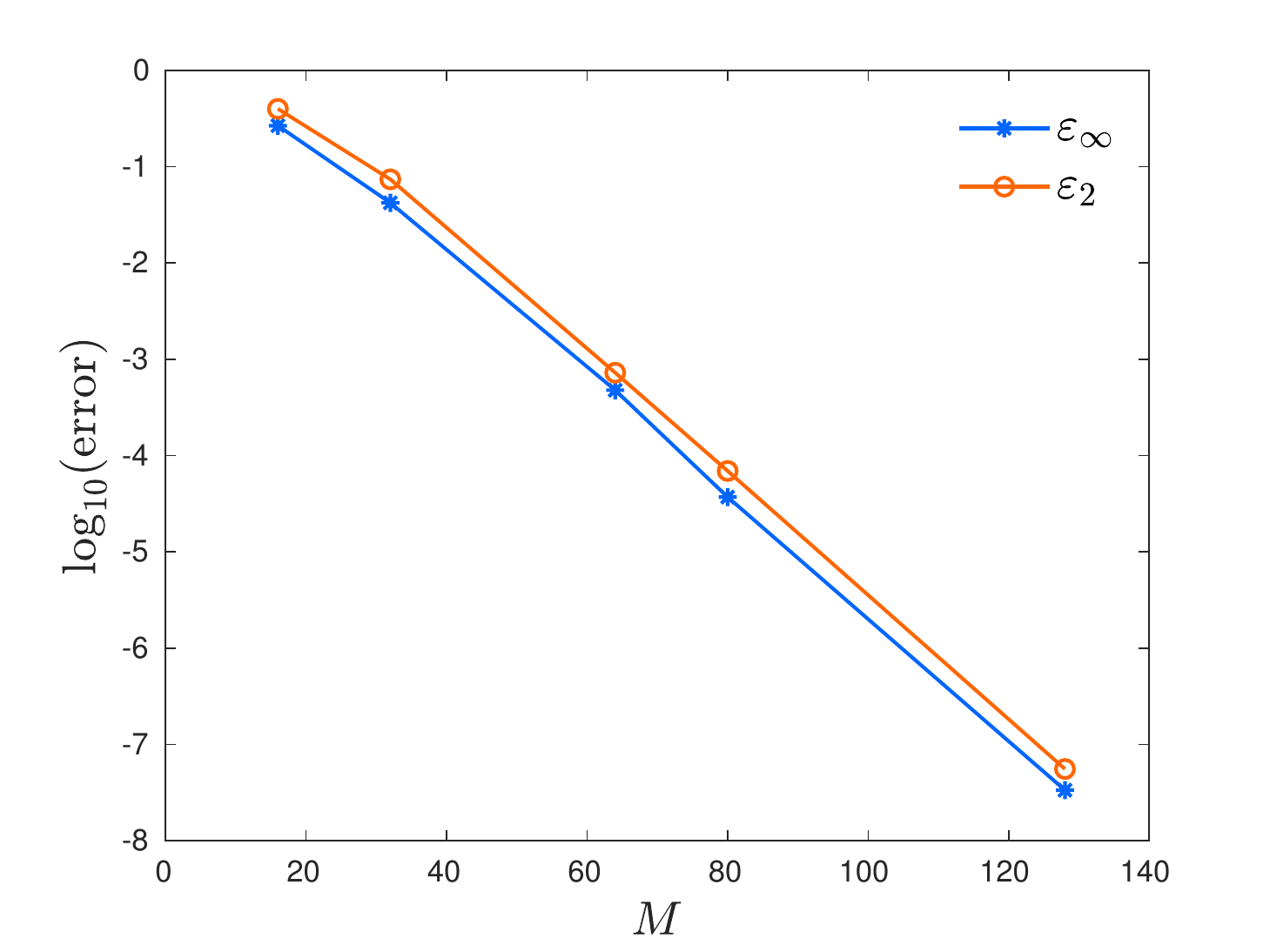}
  \caption{\small The WP system in 4-D phase space: Convergence rate with respect to $N$ (left) and $M$ (right). The spectral convergence in both $k$-space and $x$-space is clearly shown.}
  \label{fig:GBS_2d}
\end{figure}

\begin{figure}[ht!]
  \subfigure[$t=0$.]{\includegraphics[width=0.48\textwidth,height=0.33\textwidth]{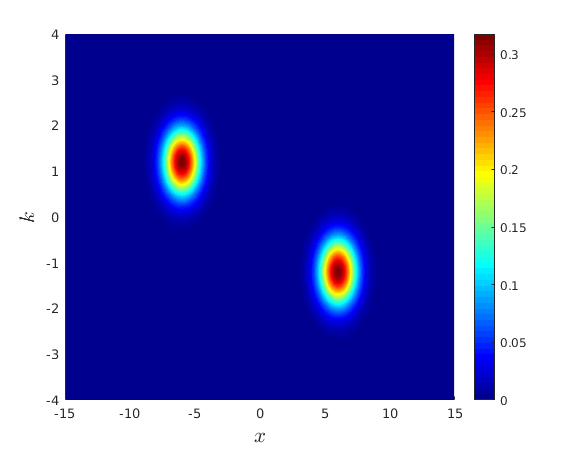}}
   \subfigure[$t=3$.]{\includegraphics[width=0.48\textwidth,height=0.33\textwidth]{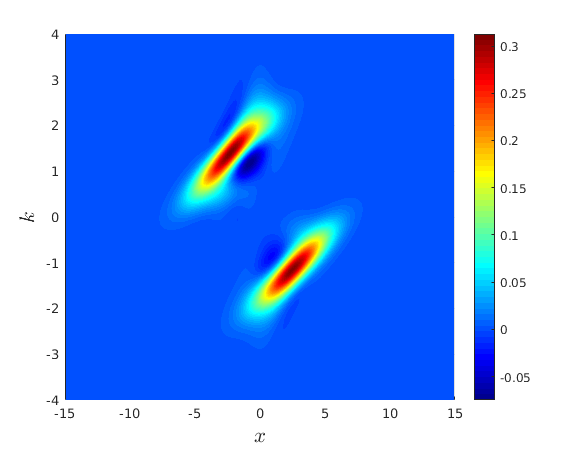}}
  
  \subfigure[$t=6$.]{ \includegraphics[width=0.48\textwidth,height=0.33\textwidth]{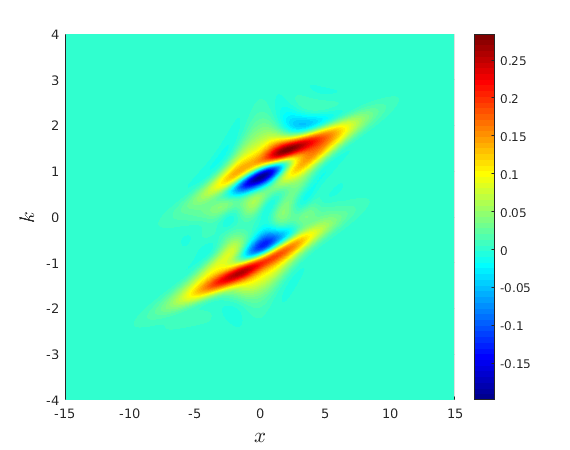}}
   \subfigure[$t=9$.]{\includegraphics[width=0.48\textwidth,height=0.33\textwidth]{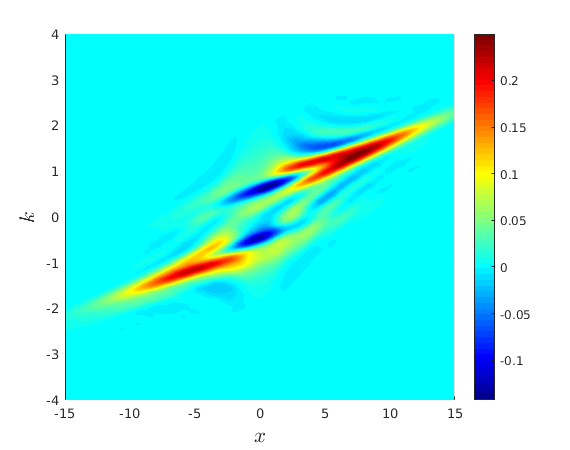}}
  \caption{\small The WP system in 4-D phase space: Reduced Wigner functions at different time instants. }
  \label{fig:GBS_2d_wigner}
\end{figure}

The numerical results are displayed in Fig.~\ref{fig:GBS_2d}, where the left (resp. right) plot shows clearly the spectral convergence with respect to $N$ (resp. $M$) while fixing $M = 200$ (resp. $N = 100$) and $\Delta t = 0.01$ fs. Moreover, we show the reduced Wigner functions of the WP system in 4-D phase space in Fig.~\ref{fig:GBS_2d_wigner}. It clearly shows that GWP crosses the barrier even when its average kinetic energy (0.72 eV) is lower than the barrier height (1.3 eV) and the Wigner functions obviously have negative values.

\section{Resonant tunneling diode}
\label{sec:rtd}

\begin{figure}[ht!]
  \centering
  \includegraphics[width=0.6\textwidth,height = 0.42\textwidth]{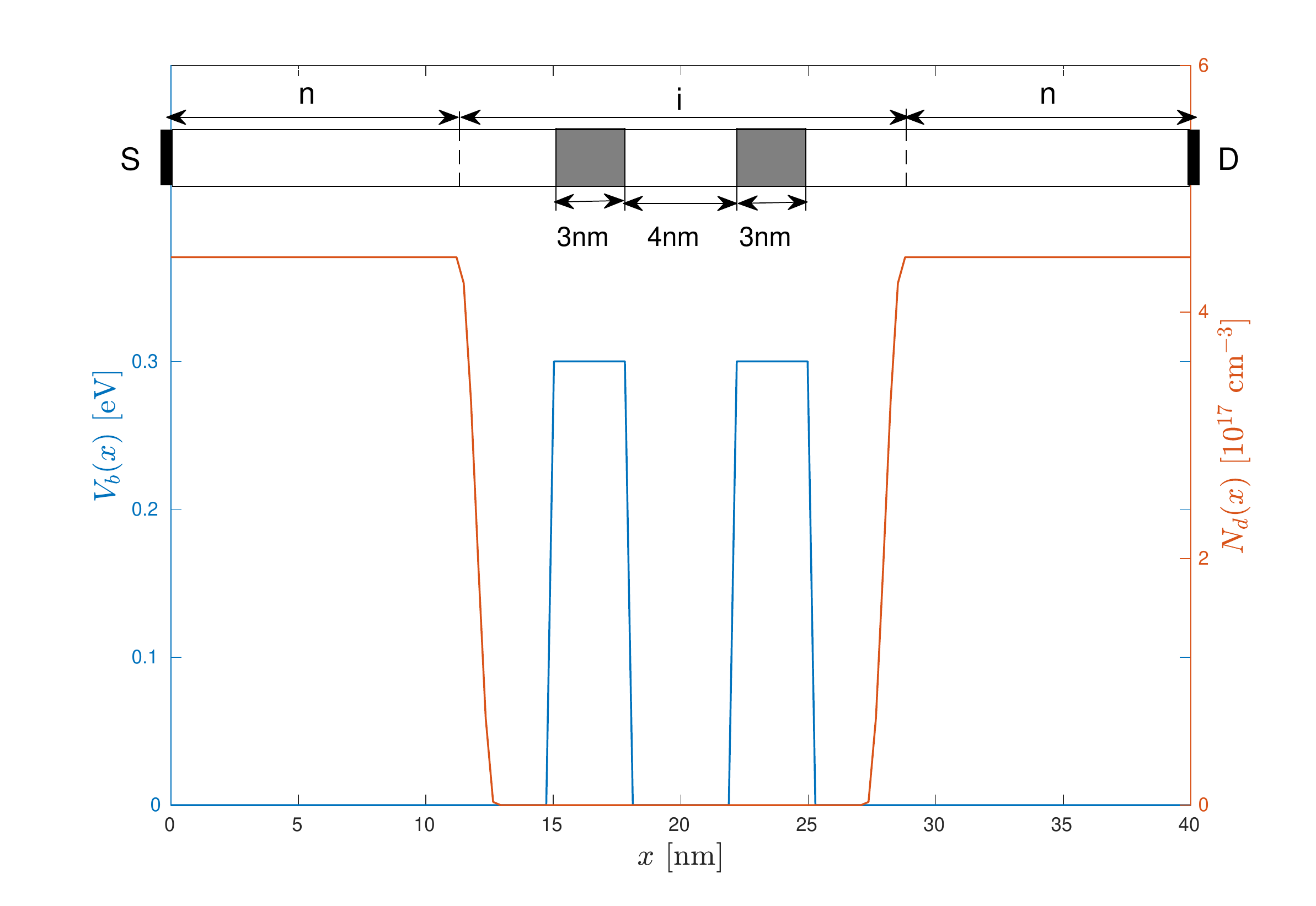}
  \caption{\small RTD structure. It consists of two n-doped contact with an intrinsic region in between which contains two barriers of 3 nm wide and height of 0.3 eV with a spacing of 4 nm. The conduction band potential $V_b(x)$ (blue solid line) and the doping density $N_D(x)=4.446\times 10^{17}~\text{cm}^{-3}$ (red dotted line) are also shown.}
  \label{fig:rtd_pot_dop}
\end{figure}

As a classical 1-D hetero-structure device with negative differential resistance,
RTD exploits resonant tunneling through double barriers as its basic mechanism.
Fig.~\ref{fig:rtd_pot_dop} gives a typical type of RTD in which 
two thin layers (gray) are sandwiched by another three layers (white) to form two energy barriers and one quantum well \cite{JiangCaiTsu2011}. In this work, we use constant effective mass $m = 0.067 m_0$ with $m_0$ being the electron mass in vacuum 
and set the length of the device to 40 nm which means the computational domain in $x$-space is $\Omega_x=[0~\text{nm},40~\text{nm}]$. The barrier region is set to 3nm, the length of the quantum well is 4 nm and the length of the contact is 10 nm. The doping profile in both contacts is depicted as the Fig.~\ref{fig:rtd_pot_dop}, where the n-parts are doped with a concentration $4.446\times 10^{17}~\text{cm}^{-3}$ and the i-part is doped intrinsically. 
The initial and boundary conditions are both taken to be fixed, and given by the equilibrium Fermi-Dirac distribution: 
\begin{align}\label{eq:Feimi-dirac1}
  f(x_l,k) &= \frac{m_0k_BT}{\pi\hbar}\log\left(1+\exp\left(\frac{\mu_L-\hbar^2 k^2/2m_0}{k_BT}\right)\right), \quad k>0,\\
  \label{eq:Feimi-dirac2}
  f(x_r,k) &= \frac{m_0k_BT}{\pi\hbar}\log\left(1+\exp\left(\frac{\mu_R-\hbar^2 k^2/2m_0}{k_BT}\right)\right), \quad k<0,
\end{align}
where $T$ is the temperature, $k_B$ is the Boltzmann constant and $\mu_L$, $\mu_R$ are the Fermi levels at the left and right contacts, respectively. The parameters are set as: $\Omega_{k} = [-\frac{5}{3}\pi~\text{nm}^{-1}, \frac53\pi~\text{nm}^{-1}]$, $m_0 = 9.10956\times 10^{-31}$ kg, $\hbar = 1.0546\times 10^{-34}~\text{J}\cdot\text{s}$ , $\varepsilon= 13.1\varepsilon_0$, $\varepsilon_0=8.85\times 10^{-12}~\text{Fm}^{-1}$, $k_BT = 2.5852\times 10^{-2}$ eV, $q_e=1.602\times 10^{-19}$ C and $\mu_L=\mu_R = 0.01$ eV. In order to get rid of possible Gibbs oscillation, a cubic interpolation (smoothing) is used over a unit near the discontinuities in the band potential $V_b(x)$ and the doping density $N_D(x)$.

We are mostly interested in the formation of steady states of RTD,
which correspond formally to the limit as $t\rightarrow +\infty$. Once the steady state is attained, the current of RTD should not appreciably vary with time any longer. To this end, we regard the numerical solution to be the steady state only when 
the difference in $L^\infty$-norm of the electron density given in Eq.~\eqref{eq:density_n} between two successive time steps is less than $10^{-5}$. Here, the time evolution is performed with a step of 0.02 fs up to the final time $t_f=500$ fs, at which the Wigner function has reached a steady state.

  
\begin{figure}[ht!]
  \centering
  \includegraphics[width=0.6\textwidth,height = 0.42\textwidth]{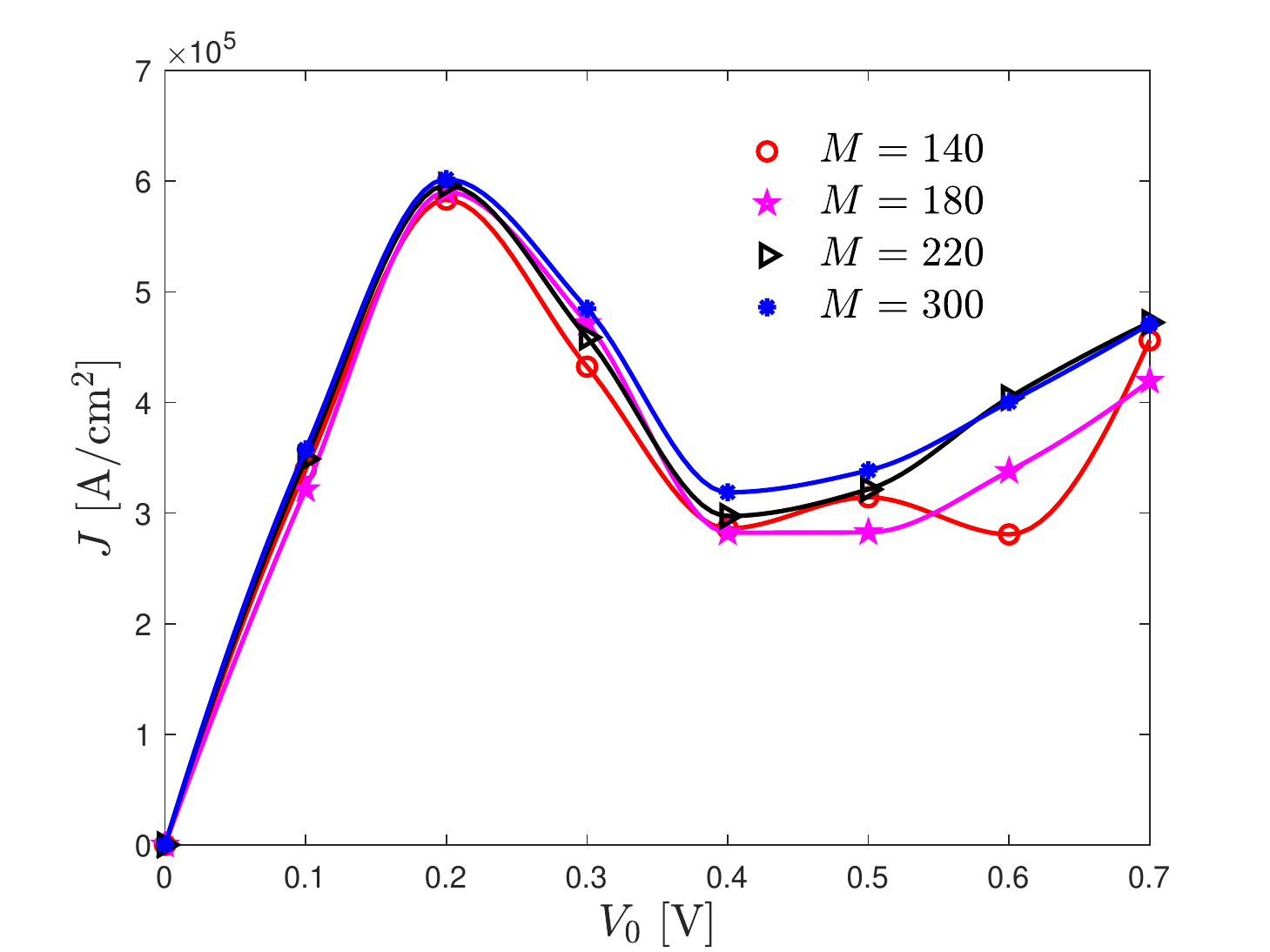}
   \caption{\small RTD: I-V curves at $t_f=500$ fs produced with four groups of $x$-grids:  $M=140, 180, 220, 300$ when fixing $N=140$ and $\Delta t = 0.02$ fs.}
  \label{fig:I-V_curve}
\end{figure}

\begin{figure}[ht!]
  \centering
   \includegraphics[width=0.48\textwidth,height = 0.35\textwidth]{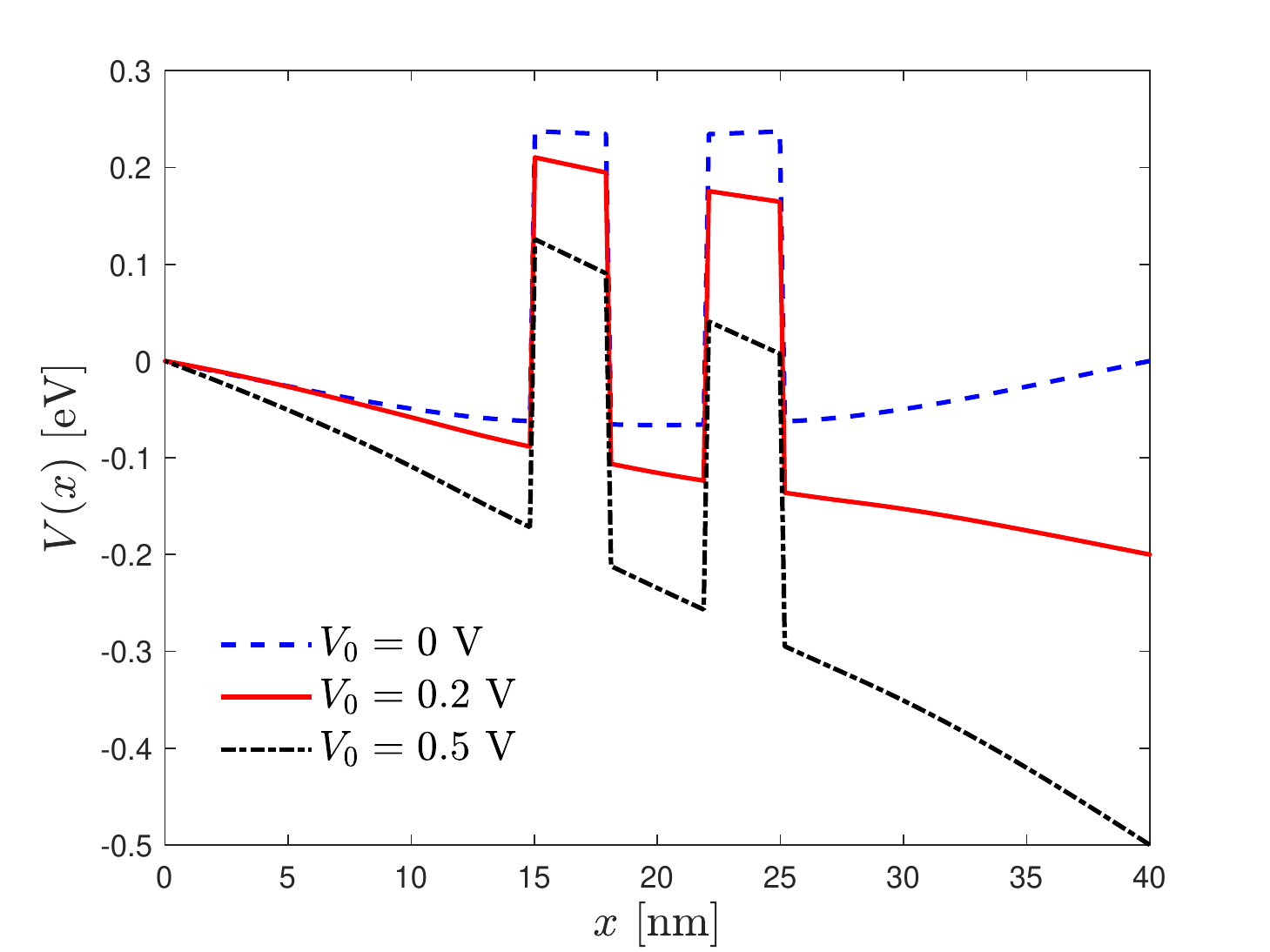}
  \includegraphics[width=0.48\textwidth,height = 0.35\textwidth]{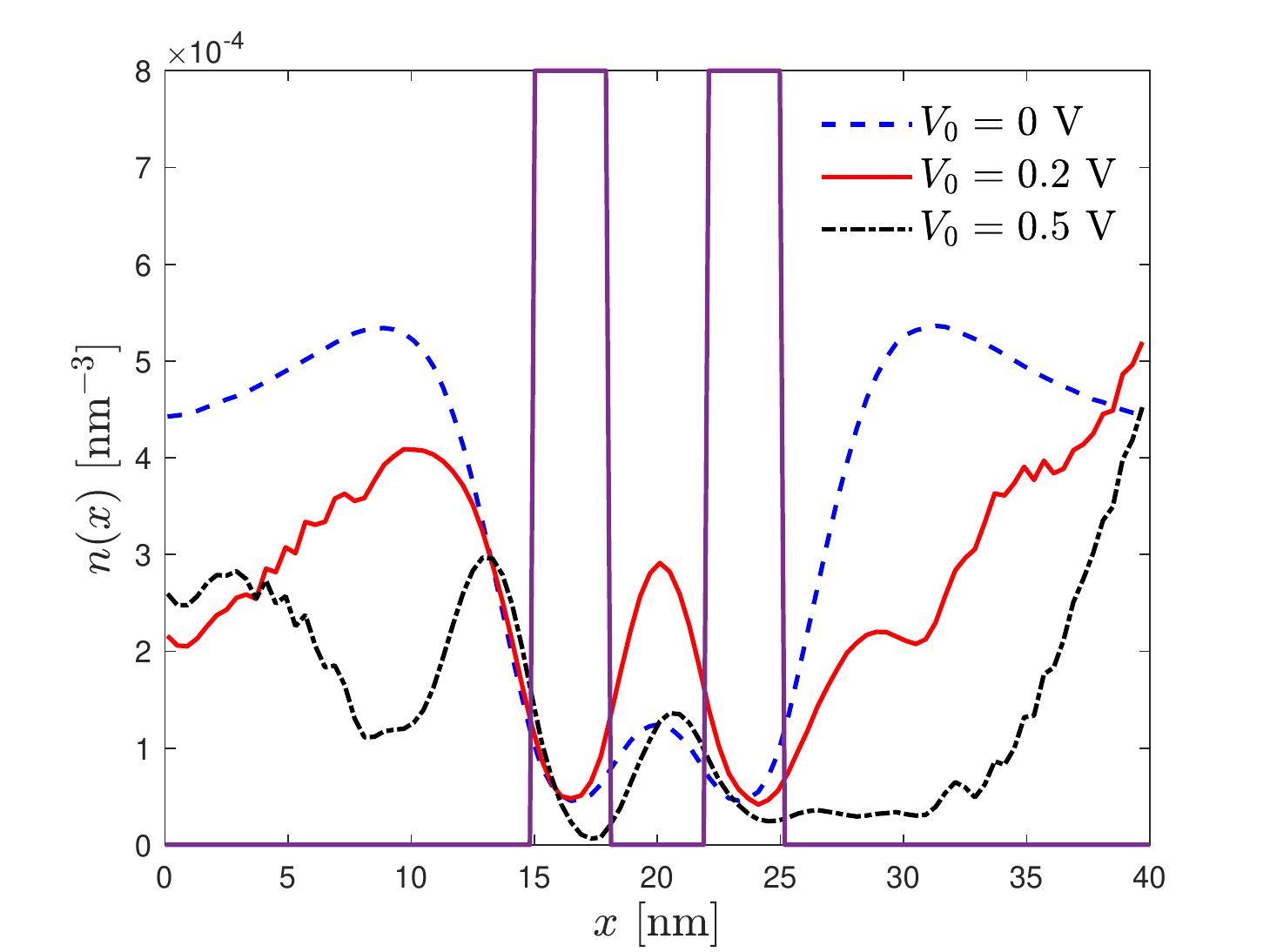}
   \caption{\small RTD: Numerically converged potential $V(x)$ (left)    and electron density $n(x)$ (right) with   bias potentials: $V_0=0$ V, $0.2$ V, $0.5$ V.}
  \label{fig:potential-density}
\end{figure}

Fig.~\ref{fig:I-V_curve} shows the I-V curves on four groups of $x$-grids:  $M=140,180,220,300$ when fixing $N=140$ and $\Delta t = 0.02$ fs. It is evident that the numerical I-V curve converges as the $x$-gird refines, 
and the results with relatively sparse $x$-grids may be unreliable. 
For example, the current density $J$ obtained with $M=140$ or $180$ shows a significant deviation  when the bias potential $V_0$ is greater than 0.4 V (see the red curve with circles).  
That is, high resolution plays a key role in producing an accurate I-V curve, which constitutes the main reason for us to develop an efficient WP solver with high accuracy. 

Second, the I-V curves in Fig.~\ref{fig:I-V_curve} show that an incoming distribution of electrons given in Eqs.~\eqref{eq:Feimi-dirac1} and \eqref{eq:Feimi-dirac2} can still generate a current flowing through the device even though under low bias like $V_0=0.1$ V. 
Simultaneously, from the converged I-V curve with $M=300$ (see the blue curve with asterisks in Fig.~\ref{fig:I-V_curve}), we are able to observe there that, the current under $V_0=0$ is zero, but it reaches a peak under $V_0 = 0.2$ V and has a valley around $V_0=0.4$ V. At even higher bias potentials, electrons surmounting the double barriers again increase the current.  However, the left plot of Fig.~\ref{fig:potential-density} shows that the height of barrier decreases with the increase of the bias $V_0$. The current peak can be reached around $V_0=0.2$ V because the resonant level in the double well is aligned with the energy of the injected electrons, which is manifested by the central peak of the electron density inside the well (see the red curve in the right plot of Fig.~\ref{fig:potential-density}).

\begin{figure}[ht!]
  \centering
  \includegraphics[width=0.32\textwidth,height = 0.26\textwidth]{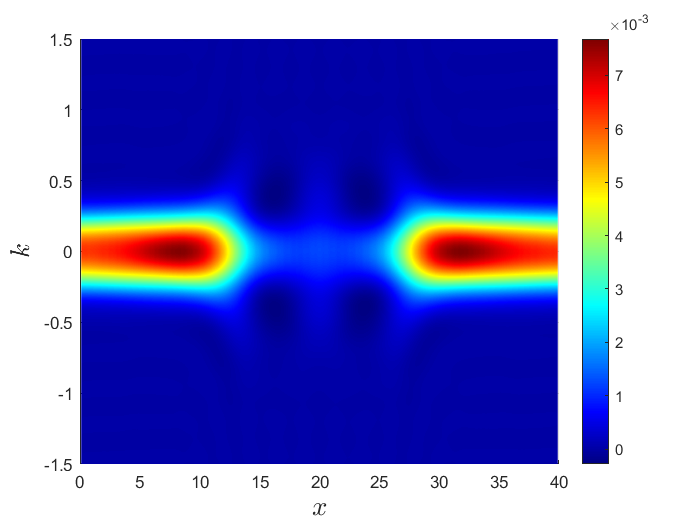}
  \includegraphics[width=0.32\textwidth,height = 0.26\textwidth]{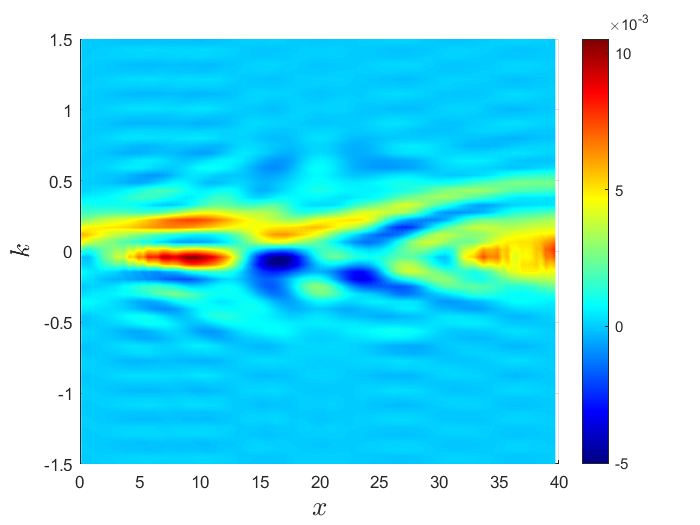}
  \includegraphics[width=0.32\textwidth,height = 0.26\textwidth]{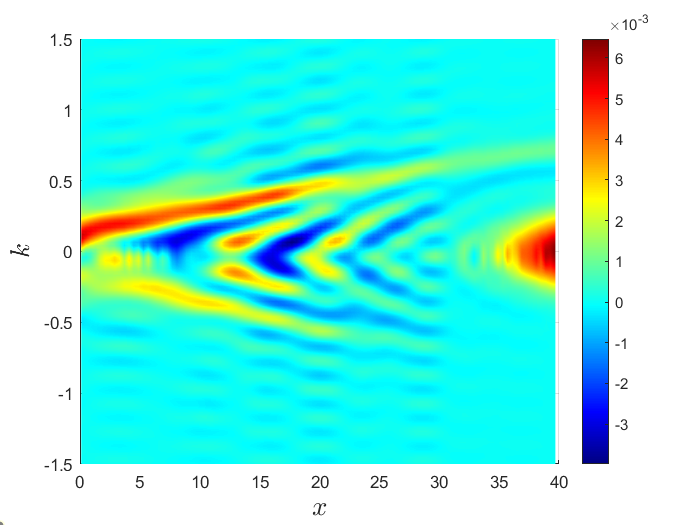}
  \caption{\small RTD: Wigner functions at the final time $t_f=500$ fs under bias potentials: $V_0 = 0$ (left), $0.2$ V (middle), and $0.5$ V (right). }
  \label{fig:rtd-wigner_k0}
\end{figure}

Finally, in Fig.~\ref{fig:rtd-wigner_k0}, we show the steady Wigner functions at the final time,
and find out that there are few electrons cross the barrier when the bias is 0, which explains that the current is almost zero under $V_0=0$. Meanwhile, it is obvious that electrons pass though the double barrier and partially reside inside the well under $V_0=0.2$ V, thereby verifying the resonance again.

\section{Double-gate MOSFET}
\label{sec:mosfet}

Fig.~\ref{fig:mosfet_device} cartoons typical structure of dgMOSFET \cite{RenVenugopalGoasguenDattaLundstrom2003, QuerliozMartinDoBournelDollfus2006}. The width of the device is assumed to be large, and the potential is invariant along $y$-direction. The silicon layer is sandwiched by two symmetric oxide layers. Source and drain are doped heavily. In this work, the size parameters of the dgMOSFET device are set as follows: the gate length $L_G$, equivalent gate oxide thickness $EOT$ and silicon channel thickness $T_{si}$ are 6 nm, 1 nm and 3 nm, respectively. The highly-doped source and drain access regions are 16 nm long. The remaining parameters are:
effective mass $m_x=m_z=0.19~m_0$, $m_y=0.98~m_0$, $m_0=9.1\times 10^{-31}$ kg, dielectric constant $\epsilon = 3.9 ~\epsilon_0$ with $\epsilon_0 = 8.85\times 10^{-12}~\text{Fm}^{-1}$, the temperature $T = 300$ K, the doping density $N_d = 5\times 10^{19}~\text{cm}^{-3}$ in the highly-doped regions.

\begin{figure}[ht!]
  \centering
  \includegraphics[width=0.6\textwidth,height = 0.42\textwidth]{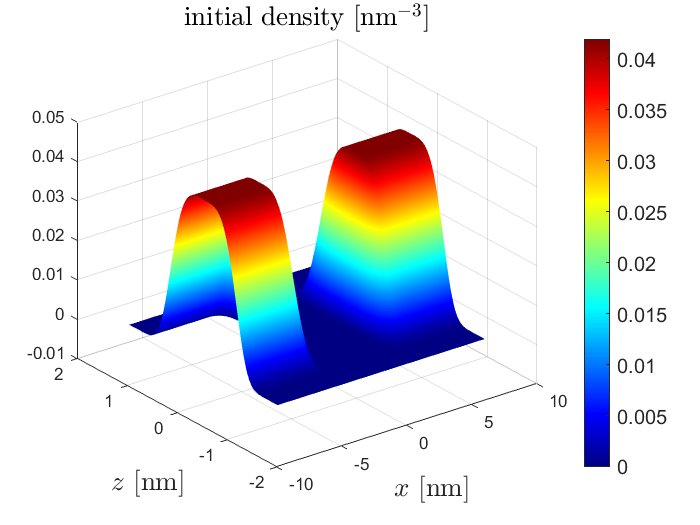}
  \caption{\small dgMOSFET: Initial density.}
  \label{fig:initial_mosfet}
\end{figure}

The electrons in the real source/drain contacts are in equilibrium characterized by a Fermi level $E_{F1}$/$E_{F2}$:
  \begin{align}
   f(x_l, \bm k, t) &= \frac{\sqrt{2m_y k_BT}}{\pi\hbar}\int_{0}^\infty \D y \frac{1}{1+\exp(y^2+\frac{\varepsilon(x_l)-E_{F1}}{k_B T})} , \quad \text{if}~ k_x > 0, \\
   f(x_r, \bm k, t) &= \frac{\sqrt{2m_y k_BT}}{\pi\hbar}\int_{0}^\infty \D y \frac{1}{1+\exp(y^2+\frac{\varepsilon(x_r)-E_{F2}}{k_B T})}, \quad \text{if}~ k_x < 0, 
  \end{align} 
 where $E_{F1} = E_{F2}= 0.0307$ eV and the total energy $\varepsilon(x)$ of the electron is
 \begin{equation}
  \varepsilon(x) = \frac{\hbar^2 k_x^2}{2m_x} + \frac{\hbar^2 k_z^2}{2m_z}.
\end{equation}
And, the initial distribution function in $z$-direction vanishes in the two oxide layers and stays constant in the semiconductor layer (see Fig.~\ref{fig:initial_mosfet}). 

We further set: $\Omega_{\bm k}=\Omega_{k_x}\times\Omega_{k_z} = [-2\pi~\text{nm}^{-1}, 2\pi~\text{nm}^{-1}]\times[-2\pi~\text{nm}^{-1}, 2\pi~\text{nm}^{-1}]$, $\Omega_{\bm x}=\Omega_x\times\Omega_z=[-8~\text{nm},8~\text{nm}] \times [-2.5~\text{nm},2.5~\text{nm}]$, the reduced Planck constant $\hbar = 1.0546\times10^{-34}~\text{J}\cdot\text{s}$, $q_e=1.602\times 10^{-19}$ C,
{gate voltage $V_{gu}=V_{gl}:=V_g$,  the time step $\Delta t = 0.025~\text{fs}$ and the final time $t_f=200~\text{fs}$, which is long enough to reach the steady state. The entire evolution takes us about 160 hours on the mesh $(N,M,\Delta t)$= (128, 140, 0.025 fs) with 28 CPUs (Intel® Xeon® @ 2.40 GHz).}

\begin{figure}[ht!]
  \centering
  \includegraphics[width=0.48\textwidth,height = 0.35\textwidth]{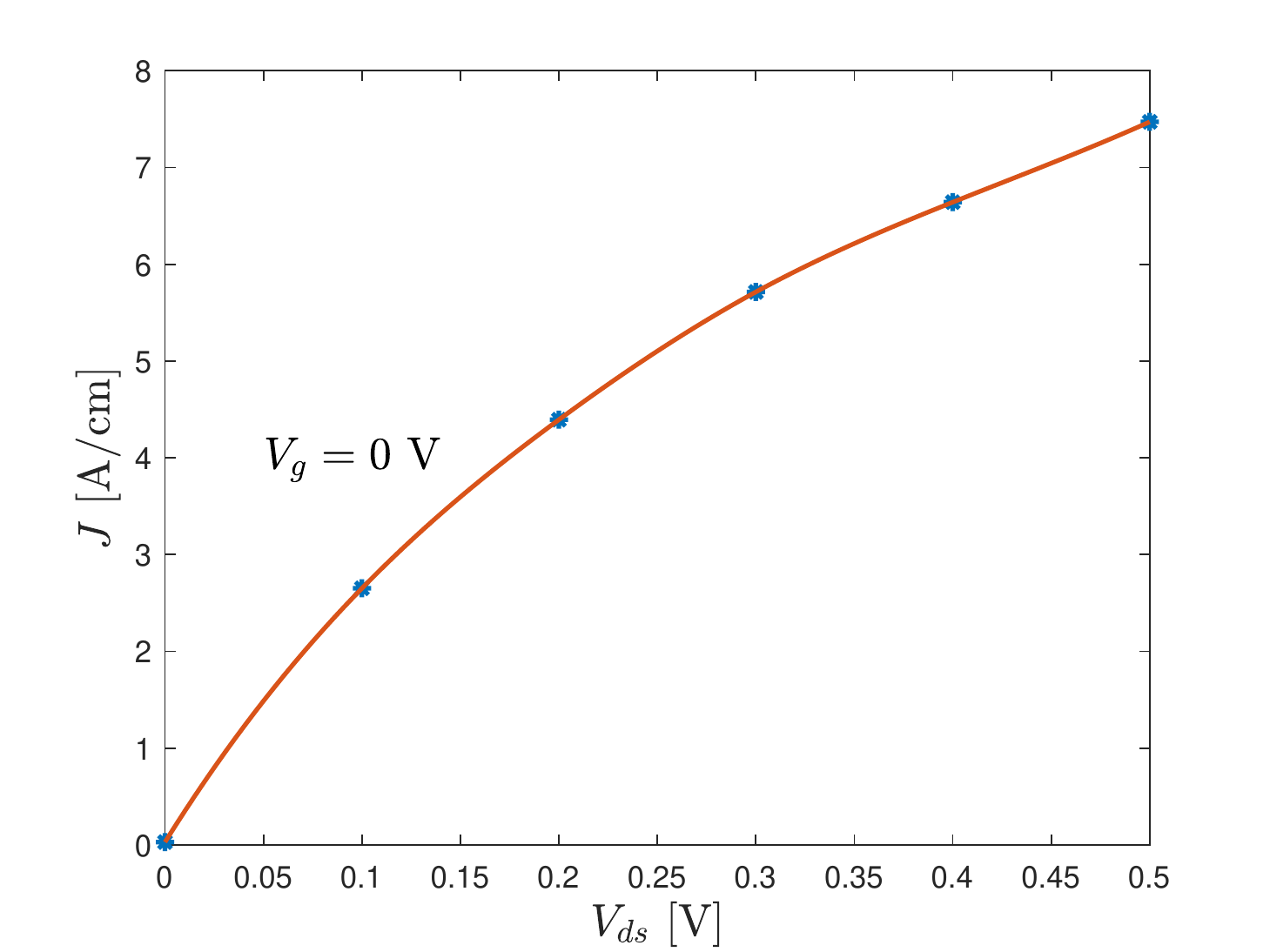}
  \includegraphics[width=0.48\textwidth,height = 0.35\textwidth]{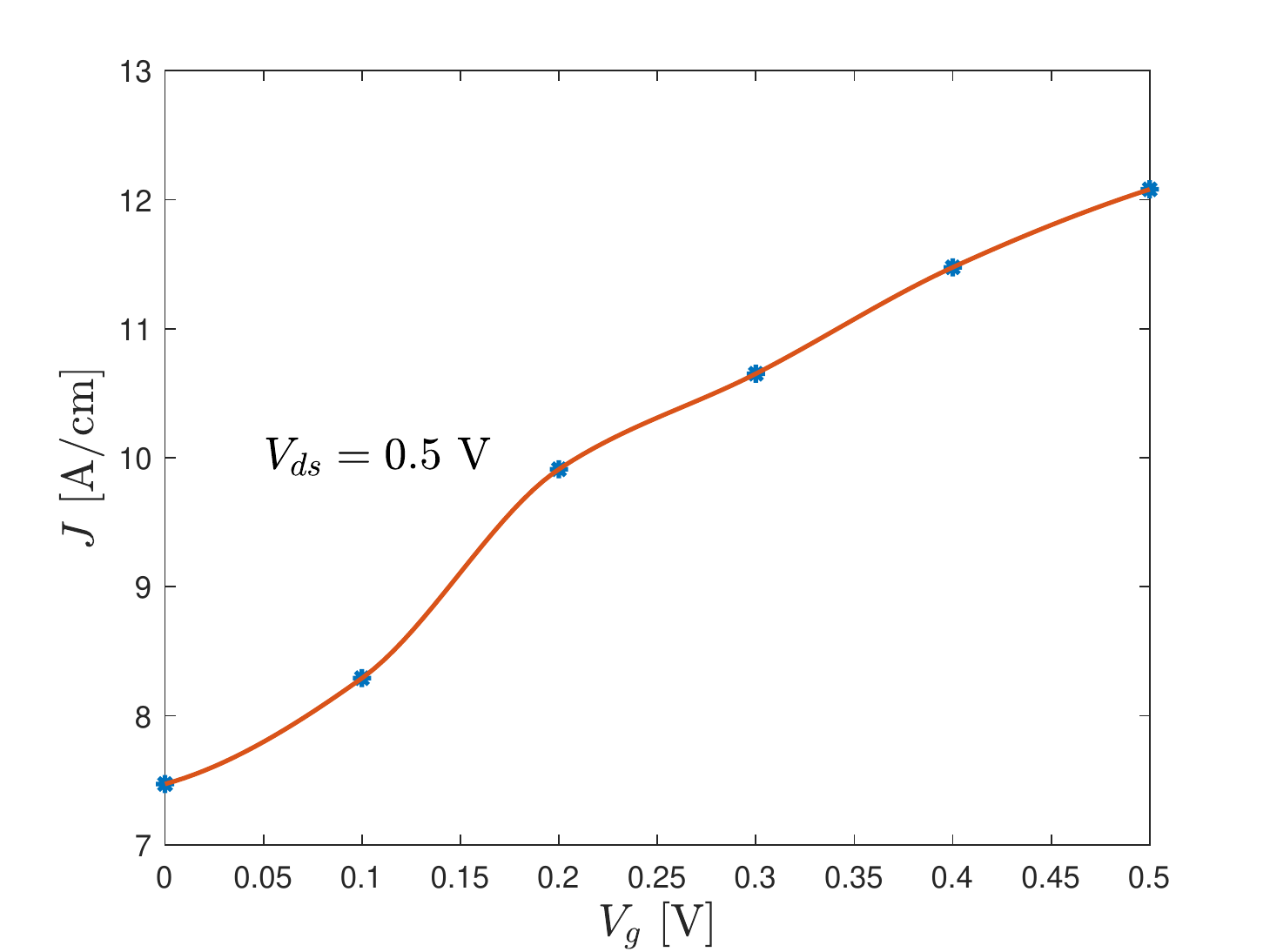}
  \caption{\small I-V curves in dgMOSFET:  Against $V_{ds}$ for $V_g=0$ V (left) and $V_{g}$ for $V_{ds}=0.5$ V (right).}
  \label{fig:iv_mosfet_k0}
\end{figure}
\begin{figure}[ht!]
  \centering
  \includegraphics[width=0.48\textwidth,height = 0.35\textwidth]{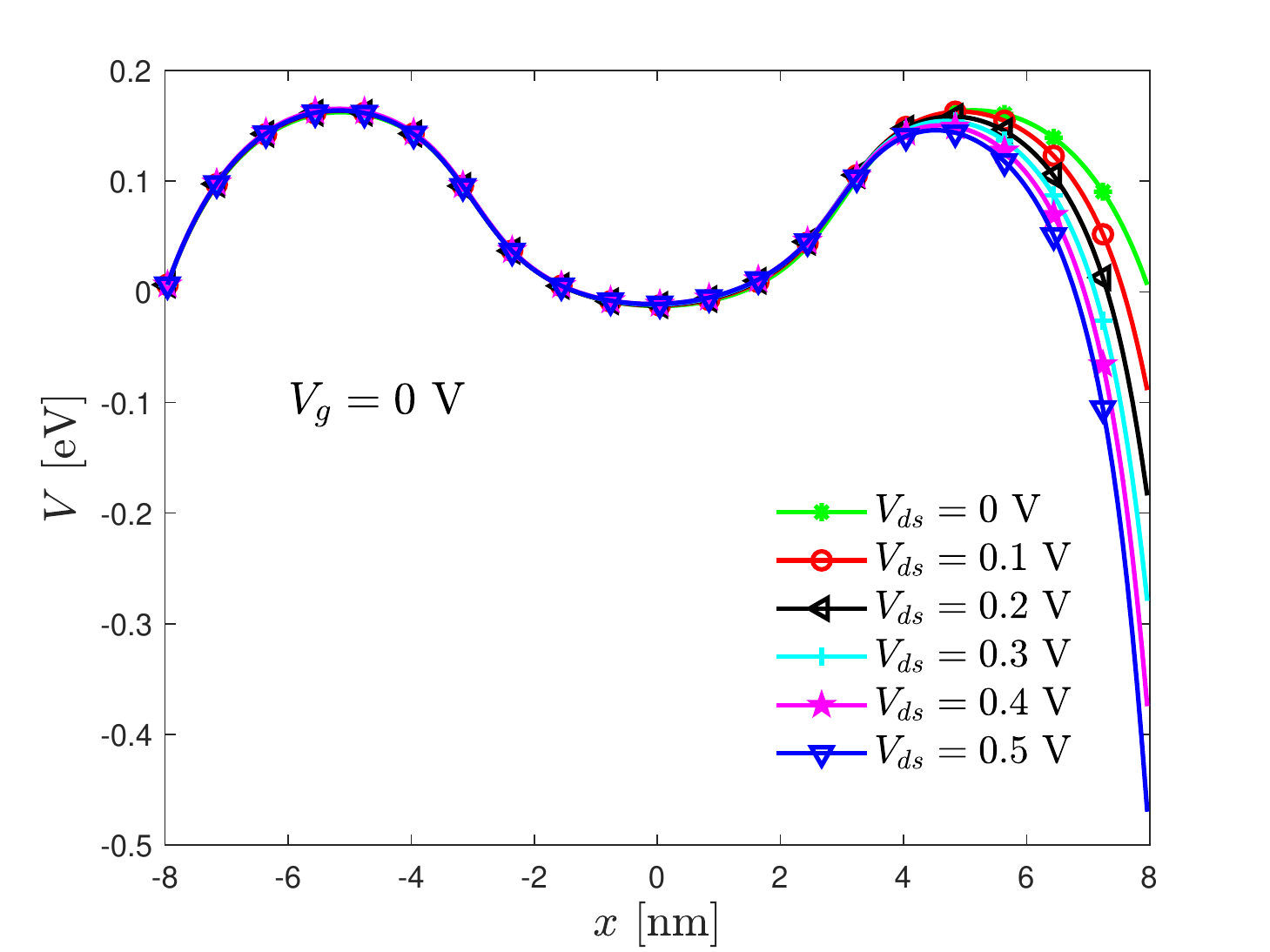}
  \includegraphics[width=0.48\textwidth,height = 0.35\textwidth]{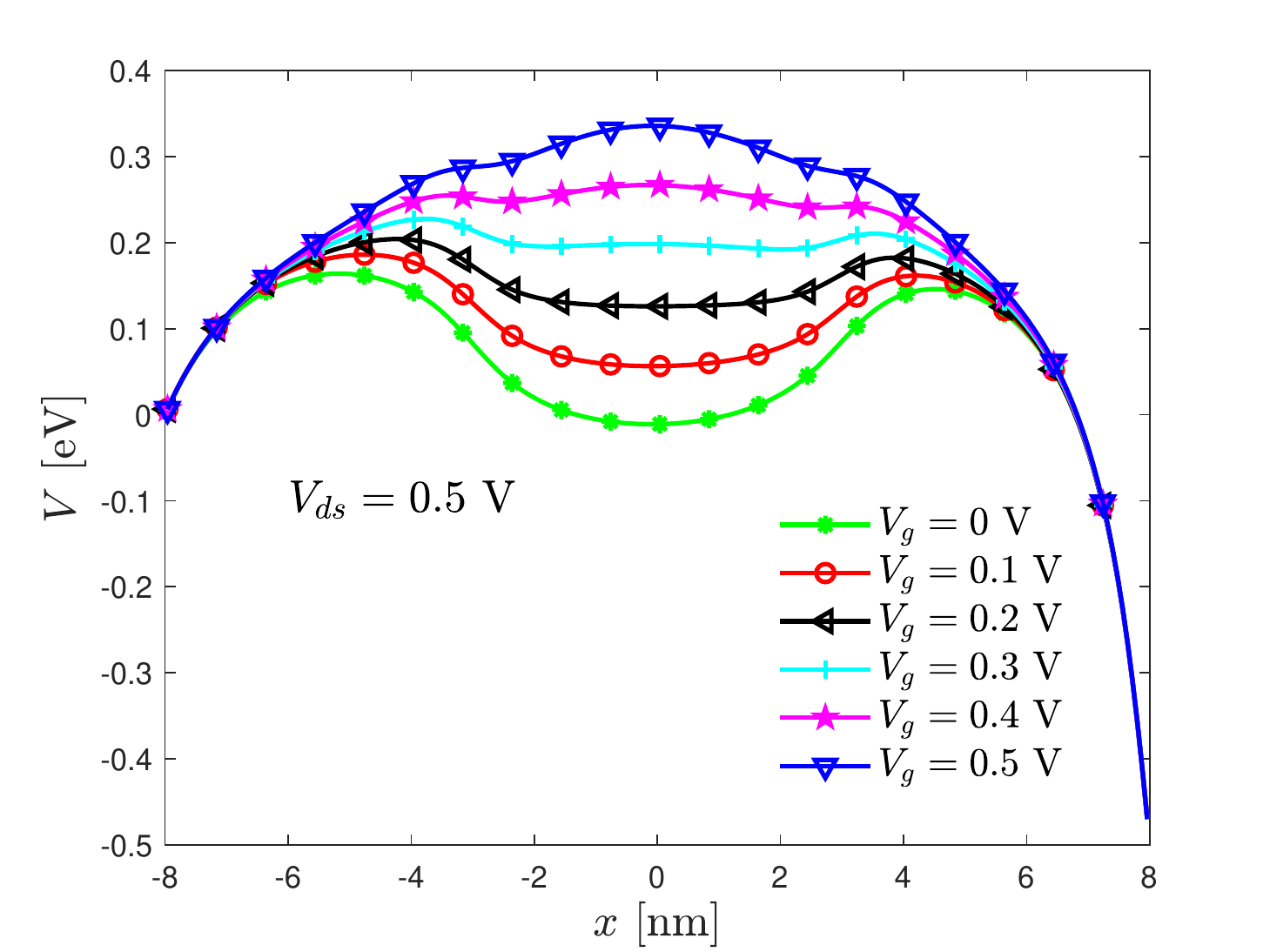}
  \caption{\small The potential along the channel in dgMOSFET:  Against $V_{ds}$ for $V_g=0$ V (left) and $V_{g}$  for $V_{ds}=0.5$ V (right).}
  \label{fig:vx_mosfet_k0}
\end{figure}

We still investigate the I-V curves but now $V$ contains the source/drain bias potential $V_{ds}$ and gate voltage $V_g$. 
Fig.~\ref{fig:iv_mosfet_k0} shows the current $J$ of steady states against $V_{ds}$ and $V_g$, where the $y$-coordinate is the integral of the current density over the contact area. 
When $V_g=0$ V is fixed, the left plot of Fig.~\ref{fig:iv_mosfet_k0} gives the I-V curves against $V_{ds}$. We find that the current increases with $V_{ds}$. 
Similarly, we plot the I-V curves against $V_g$ in the right plot of Fig.~\ref{fig:iv_mosfet_k0} for fixed $V_{ds} = 0.5$ V and observe there that the current also increases as $V_g$ increases. The trends of these two I-V curves are consistent with the results in \cite{RenVenugopalGoasguenDattaLundstrom2003}. 
We further plot the potential along the channel in Fig.~\ref{fig:vx_mosfet_k0} to explain the characteristics of the I-V curves.
The left plot of Fig.~\ref{fig:vx_mosfet_k0} shows that a larger source/drain bias $V_0$ help electrons to pass through, so the current is higher. 
The right plot of Fig.~\ref{fig:vx_mosfet_k0} displays that the increasing gate voltages $V_g$ make the the potential well shallow and eventually disappear, so the more easily the electrons pass through, the higher the current becomes.

\begin{figure}[ht!]
  \subfigure[First row: Reduced Wigner function $f(x,k_x)$.]{\includegraphics[width=0.32\textwidth,height = 0.26\textwidth]{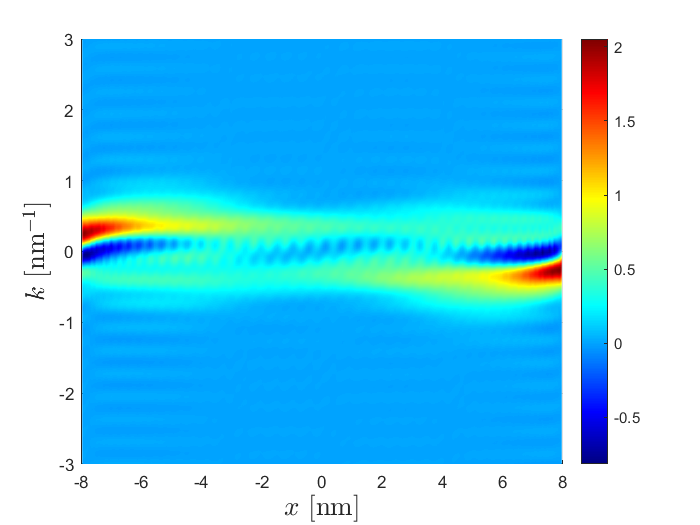}
  \includegraphics[width=0.32\textwidth,height = 0.26\textwidth]{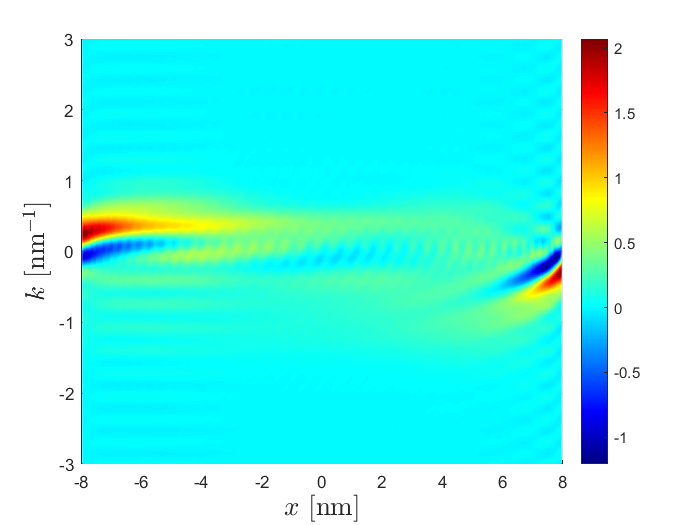}
 \includegraphics[width=0.32\textwidth,height = 0.26\textwidth]{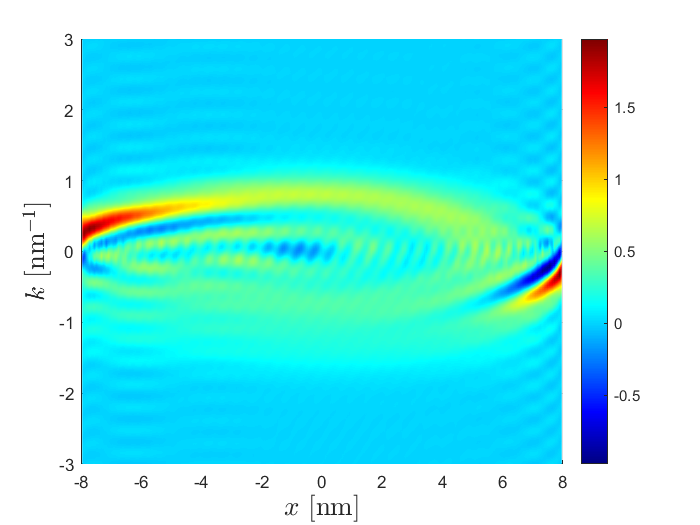}}

  \subfigure[Second row: Electric potential $V_e(x,z)$.]{\includegraphics[width=0.32\textwidth,height = 0.26\textwidth]{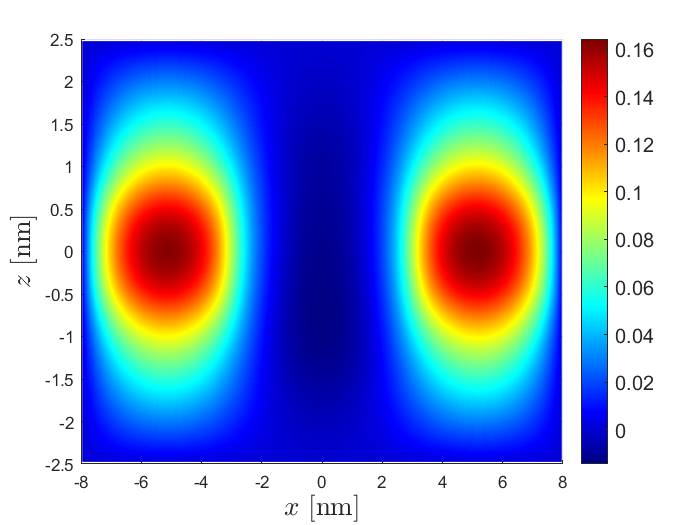}
  \includegraphics[width=0.32\textwidth,height = 0.26\textwidth]{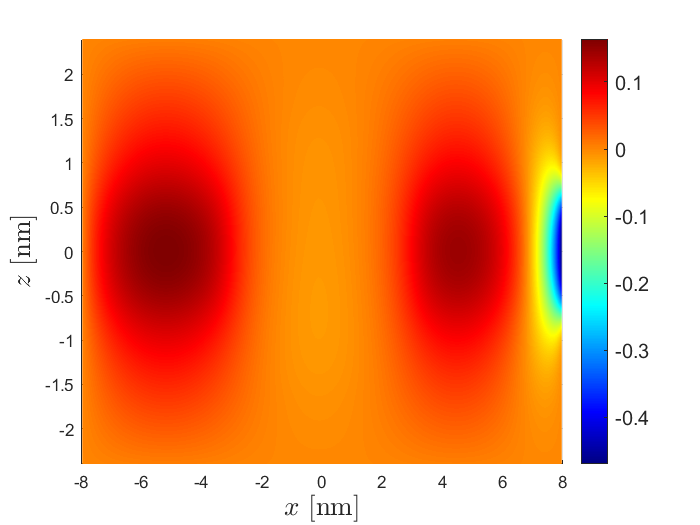}
 \includegraphics[width=0.32\textwidth,height = 0.26\textwidth]{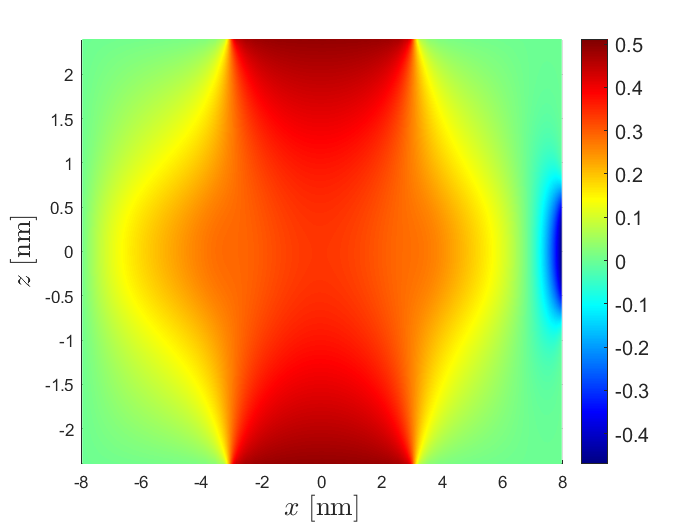}}
  \caption{\small Steady states in dgMOSFET: The reduced Wigner function $f(x,k_x)$ (first row) and electric potential $V_e(x,z)$ (second row) under $V_{ds} =0$, $V_g=0$ (first column), $V_{ds} =0.5$ V, $V_g=0$ V (second column) and $V_{ds} =0.5$ V, $V_g=0.5$ V (third column).}
  \label{fig:wigner_4d_k0}
\end{figure}

In order to give more details on the steady states, we also plot in Fig.~\ref{fig:wigner_4d_k0} the reduced Wigner function $f(x,k_x)$ (first row) and electric potential $V_e(x,z)$ (second row) under different $V_{ds}$ and $V_g$.
The first row of Fig.~\ref{fig:wigner_4d_k0} shows that electrons flow more easily from the left to the right as the bias $V_{ds}$ increases. And at the  same time, increasing the gate voltage $V_g$ is also beneficial to the flow of electrons.
The second row of Fig.~\ref{fig:wigner_4d_k0} displays that the doping forms a barrier when $V_g=0$, so that the intermediate channel forms a potential well, but the well depth decreases as $V_g$ increases.

\section{Conclusion}
\label{sec:conclude}

In this paper, we made the first attempt to solve the Winger-Poisson system in 4-D phase space with high accuracy,  
and succeeded to
develop steady states and to obtain numerically converged I-V curves from reliable long-time simulations. 
We believe that the proposed high-resolution solver may provide more reference solutions to benchmark the stochastic algorithms which have
recently attracted a lot of attention due to its simplicity as well as its satisfactory
scaling on parallel high-performance machines \cite{MuscatoWagner2016,XiongShao2019,bk:NedjalkovDimovSelberherr2021,BenamBallicchiaWeinbubSelberherrNedjalkov2021}.


\section*{Acknowledgement}

This work was supported by the National Key R~\&~D Program of China (No. 2020AAA0105200), the National Natural Science Foundation of China (Nos. 11822102 and 12171035) and China Postdoctoral Science Foundation (No. 2021M690467).
ZC thanks Institute of Applied Physics and Computational Mathematics in Beijing for providing high-performance computing platform. 
SS acknowledges Beijing Academy of Artificial Intelligence (BAAI) and the computational resource provided by High-performance Computing Platform of Peking University.
The authors are grateful to the useful discussions with Yunfeng Xiong,
as well as to the handling editor and the referees for their patience and very valuable suggestions.

\section*{Appendix}

We have 
{\scriptsize
\begin{equation}\label{eq:P}
    P = \left(
    \begin{array}{cccccccc}
      1 & 0 & 1 & 0 & \cdots & 1 & 0 & 1 \\
      0 & 1 & 0 & 1 & \cdots & 0 & 1 & 0 \\
      0 & 0 & \frac{2^3}{2} & 0 & \cdots & \frac{(N-2)^3}{2} & 0 & \frac{N^3}{2}\\
      0 & 0 & 0 & 3(3^2-1) & \cdots & 0 &  (N-1)((N-1)^2-1) & 0 \\
      \vdots&\vdots&\vdots&\vdots&\vdots&\vdots&\vdots&\vdots\\
      0 & 0 & 0 & 0 & \cdots & 0 & (N-1)((N-1)^2-(N-3)^2) & 0\\
      0 & 0 & 0 & 0 & \cdots & 0 & 0 & N(N^2-(N-2)^2)\\
    \end{array} \right)
\end{equation}}

and

{\scriptsize
\begin{equation} \label{eq:I_tilde}
 \tilde{I} = \left(
    \begin{array}{cccccc}
      0 & & & &  \\
      0 & &\ddots &  & &\\
      1 & & \ddots & \ddots  & &\\
        & & \ddots &\ddots  & \ddots \\
        & & & 1 & 0 & 0\\
    \end{array} \right).
\end{equation}}

Then, the linear matrix system Eq.~\eqref{eq:matrix_mix} equals to the following block upper triangular equation 

{\scriptsize
\begin{equation}\label{eq:matrix_mix_2}
  \left(
  \begin{array}{cccccccc}
    I & 0 & I & 0 & \cdots & I & 0 & I \\
    0 & I & 0 & I & \cdots & 0 & I & 0 \\
    0 & 0 & \tilde{A}_{02}& 0 &\cdots & \tilde{A}_{0,M-2}& 0 & \tilde{A}_{0M}\\
    0 & 0 & 0 & \tilde{A}_{13} & \cdots & 0 & \tilde{A}_{1,M-1} & 0\\
    \vdots&\vdots&\vdots&\vdots&\vdots&\vdots&\vdots&\vdots\\
    0 & 0 & 0 & 0 & \cdots  & 0 & \tilde{A}_{M-3,M-1}  & 0\\
    0 & 0 & 0 & 0 & \cdots & 0 & 0 & \tilde{A}_{M-2,M} \\
  \end{array}\right)
  \left(
  \begin{array}{c}
    X_0\\
    X_1\\
    X_2\\
    X_3\\
    \vdots\\
    X_{M-1}\\
    X_M\\
  \end{array}\right) =
 \left(
  \begin{array}{c}
    G_0\\
    G_1\\
    G_2\\
    G_3\\
    \vdots\\
    G_{M-1}\\
    G_M\\
  \end{array}\right),
\end{equation}
}

where

{\small
\begin{align*}
  &\tilde{A}_{0i} = A_{0j}-P,~~ j = 2,4,\ldots, M, \\
  &\tilde{A}_{1i} = A_{1j}-P, ~~ j = 3,5,\ldots, M-1, \\
  &\tilde{A}_{ij} = A_{ij} - P\tilde{A}_{i-2,i}^{-1}\tilde{A}_{i-2,j}, ~~i =2,\ldots, M-2,~~
    j = i+2,\ldots, M,\\
  & G_0 = \tilde{F}_0, \quad G_1 = \tilde{F}_1,\\
  & G_{2} = \tilde{F}_2 - PG_0, \quad  G_{3} = \tilde{F}_3 - PG_1,\\
  & G_{i} = \tilde{F}_i - P\tilde{A}_{i-2,i}^{-1}G_{i-2}, ~~ i = 4,\ldots, M.
\end{align*}
}

Then we get the solution of Eq.~\eqref{eq:matrix_mix_2} as follows
{\small
\begin{align*}
  & X_M = (\tilde{A}_{M-2,M})^{-1}G_M, \\
  & X_{M-1} = (\tilde{A}_{M-3,M-1})^{-1}G_{M-1}, \\
  & X_{i} = (\tilde{A}_{i-2,i})^{-1}\left[G_{i} - \sum_{j=i+2:2:N}\tilde{A}_{i-2,j}X_{j}\right], ~~ i = M-2,\ldots, 2,\\
  & X_{1} = G_1 - \sum_{j=3:2:M-1}X_{j}, \\
  & X_{0} = G_0 -\sum_{j=2:M} X_{j}.
\end{align*}}

\end{document}